\documentstyle[12pt]{article}
\begin{document}
\def\e{\mbox{e}}
\def\sgn{{\rm sgn}}
\def\gsim{\;
\raise0.3ex\hbox{$>$\kern-0.75em\raise-1.1ex\hbox{$\sim$}}\;
}
\def\lsim{\;
\raise0.3ex\hbox{$<$\kern-0.75em\raise-1.1ex\hbox{$\sim$}}\;
}
\def\MeV{\rm MeV}
\def\eV{\rm eV}
\thispagestyle{empty}

\vskip 0.3cm

%\documentstyle[12pt]{article}
%\topmargin -1cm
%\textwidth 6in
%\textheight 8.5in
%\evensidemargin  0cm
%\input psfig

%\begin{document}
%\def\e{\mbox{e}}
%\def\sgn{{\rm sgn}}
%\def\gsim{\;
%\raise0.3ex\hbox{$>$\kern-0.75em\raise-1.1ex\hbox{$\sim$}}\;
%}
%\def\lsim{\;
%\raise0.3ex\hbox{$<$\kern-0.75em\raise-1.1ex\hbox{$\sim$}}\;
%}
%\def\MeV{\rm MeV}
%\def\eV{\rm eV}
%\thispagestyle{empty}

\vskip 0.3cm
%\documentstyle[prd,aps,eqsecnum,oldlfont]{revtex}
%\begin{document}
%\draft
%\newcommand{\be }{\begin{equation}}
%\newcommand{\bea }{\begin{eqnarray}}
%\newcommand{\en }{\end{equation}}
%\newcommand{\ena }{\end{eqnarray}}

%\def\e{\mbox{e}}
%\def\sgn{{\rm sgn}}
%\def\gsim{\;
%\raise0.3ex\hbox{$>$\kern-0.75em\raise-1.1ex\hbox{$\sim$}}\;
%}
%\def\lsim{\;
%\raise0.3ex\hbox{$<$\kern-0.75em\raise-1.1ex\hbox{$\sim$}}\;
%}
%\def\MeV{\rm MeV}
%\def\eV{\rm eV}

\begin{center}
\hfill astro-ph/9812149\\
\hfill{\today}\\

\LARGE

{\bf MHD waves as a source of matter density
fluctuations within solar interior}

\end{center}

\large
\vskip 0.3cm
\begin{center}
{\bf N.S. Dzhalilov}~\footnote{E-mail:
namig@izmiran.rssi.ru} and {\bf V.B.
Semikoz}~\footnote{E-mail:  semikoz@orc.ru}
\end{center}
\normalsize %
\begin{center} {\it  The Institute of the Terrestrial Magnetism,
the Ionosphere and Radio Wave Propagation of the Russian Academy of
Sciences\\ IZMIRAN,Troitsk, Moscow region, 142092, Russia} \end{center}
\begin{abstract}
It is shown that in the presence of a background magnetic field
within solar interior a cavity
for low frequency MHD eigen modes (with periods 1-10 days)
near equatorial plane can arise. The lower boundary of the cavity
coincides with the center of the Sun while the upper one corresponds to
the Alfven resonant layer where high accumulation of wave energy takes
place. The localization $z_s$ and the width of the Alfven resonance layer
$\Delta z_s$ are determined by: (i) the node number of eigen modes $n = 1,
2,...$, (ii) by the angle $\alpha$ of oblique propagation of waves with
respect to the magnetic field {\bf B}, and (iii) by a low
magnitude of the background magnetic field itself, $B_0 =
1-100~G$.  The amplitude of eigen oscillations in a resonant layer
determines the density fluctuation value $\delta \rho/\rho$ that is
restricted through the imaginary part of eigen frequences.

For large node numbers $n\gg 1$ there appear many narrow
resonant layers where a neutrino propagates through a large
density fluctuation $\delta \rho/\rho$ with the oscillation
length that is much bigger than the width of a resonant layer,
$l_{osc} = 4\pi E/\Delta m^2\gg \Delta z_s(n)$.
It is shown that neutrino crosses many such bumps on the exponential
background profile that motivates to consider these MHD waves as a
plausible matter noise for the MSW solution to the Solar Neutrino
Problem (SNP).

\end{abstract}

\vskip 0.3cm
PACS codes: 13.10.+q; 13.15.-f; 13.40.Fn; 14.60.Gh; 96.60.Kx.\\
{\bf Key words}: neutrino, magnetic fields, magnetohydrodynamics,
wave resonant absorption, cavity.

\newpage

\section{Introduction}
\vskip 0.3cm

    The existence of the solar neutrino problem is now established from
five underground experiments \cite{Lande,Fukuda,Gavrin,Kirsten,Suzuki}
more or less independently of any details of the Standard Solar Model
(SSM) \cite{Bahcall1}.

At the same time it becomes clear \cite{Fiorentini} that any modifying
SSM astrophysical solution contradicts the reconciled data of the
Homestake \cite{Lande} and the Kamiokande \cite{Fukuda} experiments.
Moreover, Helioseismology data are in a good agreement\cite{Berezinsky}
with astrophysical parameter values given by SSM and relevant to all
Particle Physics solutions: (i) neutrino vacuum oscillations
\cite{vacuum}, (ii) resonant neutrino oscillations in medium or the
Mikheev-Smirnov-Wolfenstein (MSW) solution \cite{MSW} and (iii) the
Resonant Spin-Flavor Precession (RSFP) in medium\cite{RSFP}.

The most popular Small Mixing Angle (SMA) MSW solution\cite{MSW} depends on
minimal number of unknown fundamental parameters (neutrino mass
difference $\Delta m^2$ and mixing $\sin^22\theta\ll 1$) and relies on
resonant neutrino conversions ($\nu_{eL}\to \nu_{\mu L}$) in matter
with the smooth density profile $\rho_0 = \rho_{00}\exp (- r/H)$ where $H =
0.1R_{\odot}$ is the height scale in SSM \cite{Bahcall}. Any change of this
profile generally may be crucial for allowed fundamental parameter region
(see below) while standard Helioseismology corrections to SSM give density
fluctuations deep in the solar interior at the level $\delta
\rho/\rho_0\lsim
1$\%\cite{Dziembowski} that occurs too small to influence neutrino
oscillations.

Another scenario (RSFP \cite{RSFP})
includes the additional parameter $\mu B_0$ where $B_0$ is an
extrapolation of a regular large-scale magnetic field
(as seen on the solar surface)
down to the bottom of the convective zone and $\mu$  is an unknown
transition
Majorana (Dirac) neutrino magnetic moment.
The existence of such large-scale magnetic fields $B_0$
under the bottom of the convective zone and even in the solar core is
currently discussed (see, for instance, \cite{Parker}), and it is not
excluded that these fields can lead to the MHD wave generation in the solar
interior relevant to the MSW-resonance region for neutrino oscillations
since density perturbations $\delta \rho (\vec{r},t)/\rho_0$ ( as
well as the magnetic field ones, $\vec{b}(\vec{r},t)/B_0$,  and the speed
fluctuations $\vec{v}(\vec{r},t)$) are evolved in the presence of a
central field $B_0$.

Therefore, the main goal of present work is the search of possible magnetic
field
enhancement of the density perturbations $\delta \rho/\rho_0$ resulting in
an effective non-direct neutrino interaction with an external magnetic
field $B_0$ {\it without any neutrino magnetic moment}. \footnote{Another
example of neutrino interaction with magnetic field and also without
neutrino magnetic moment was found in \cite{SV} where the mean axial
vector potential $V_A\sim G_F\langle
\bar{\psi}_e\gamma_z\gamma_5\psi\rangle\sim B_0$ for an electron gas
polarized in a strong magnetic field {\bf B} = $(0,0,B_0)$ \cite{NSSV}
was calculated. This polarization leads to anisotropic
neutrino emission through MSW oscillations studied in \cite{KS} in
order to explain kick of a neutron star. For the solar case the axial
vector
potential is not efficient since the solar magnetic field is too small to
polarize the Boltzman electron gas. Even at the bottom of the convective
zone with the strongest magnetic field value $B_0\sim 10^5~G$ the
paramagnetic contribution is $\mu_BB_0\lsim 10^{-3}~eV$ only and the
axial vector potential $V_A$ is much less than the vector potential
$V$, $V_A\sim V\mu_BB_0/T\ll V$ where the standard MSW
vector potential is defined as
$V= G_F\sqrt{2}\rho Y_e/m_p$ for $\nu_e\to \nu_{\mu}$
i.e. it is proportional to the mean matter density $\rho$. Here
$G_F$  is the Fermi constant, $\mu_B$ is the Bohr magneton, $T$
is the temperature within the Sun, $Y_e$ is the electron abundance.}

Here we  consider
only  very modest values of large-scale background magnetic fields in
central
region of the Sun, $B_0\sim 1~-10~-100~G$, that is not in contradiction
with a primeval origin of such central field in the stars of main sequence
like the
Sun and with non-observation of traces of such small (dipole) fields in the
solar photosphere. This is in contrast with the  visible traces of more
stronger
toroidal (and poloidal) fields generated at the bottom of the convective
zone.

In particular, the most interesting region $B_0\lsim 10~G$ where MHD
density perturbations are similar to the intense matter noise
(see below) obeys the constraint on the central field $B_0\lsim 30~G$ found
recently \cite{primeval} from the assumption of primeval origin
of solar magnetic fields which are evolved via dynamo mechanism.

There were some previous attempts to modify the
SSM density profile $\rho_0$ \cite{Bahcall} entering the neutrino
evolution equation in the MSW solution to SNP.
The influence of periodic matter density perturbations above the average
density $\rho_0$, i.e.  \begin{equation} \rho (r) = \rho_0[1 + h\sin
(\gamma r)]~, \label{simple} \end{equation} on resonant neutrino
conversion was investigated in Ref. \cite{Krastev}. In this case
the parametric
resonance effect appears
when the fixed frequency ($\gamma$) of the perturbation is close to the
neutrino oscillation frequency and a large amplitude value
($h\sim 0.1-0.2$) is assumed. There are also a number of works which
address to similar effects by different approaches \cite{Petcov}.

One of disadvantages of models like Eq. (\ref{simple}) is our unawareness
of values of model parameters $\gamma$, $h$ as well as of  physical origin
of these perturbations.

Another approach was developed in Ref.\cite{Hiroshi} where authors
assumed the presence of a white noise of matter density ($\langle
\xi\rangle =\langle \delta \rho/\rho_0\rangle = 0$) added to $\rho_0$
instead of the second term in Eq.  (\ref{simple}).  The number of free
parameters (the noise amplitude $\sqrt{\langle \xi^2\rangle}$ and the
correlation length $L_0$ for random $\delta$-correlated domains)
remains the same one (=2) while the presence of white noise in the Sun is
doubtless since there are many mechanisms to enhance random density
perturbations.

However, Bamert et al (c.f.,for instance, \cite{Burgees}) stated that in
analysis of matter noise influence the MSW solution to SNP  given in
\cite{Hiroshi} (as well as in their own first work and in
\cite{Balantekin}) what has been missing so far is {\it a plausible
source for such $\delta$-correlated fluctuations in the vicinity of the
MSW resonant point}.

Assuming that a certain superposition of g-modes can be a source of such
fluctuations deep in the solar interior, they showed that even for
large amplitudes of some g-modes
(or large $\delta \rho/\rho\sim 4$\%) for a large size of density
fluctuation cell,
$L_0\gsim 3000~km$, the direct numerical
integration of evolution equation leads to  standard MSW result.

Comparing this value with the neutrino oscillation length at the MSW
resonance,
$l_{osc}^{res} = l_v/\sin 2\theta = 250~km(E/MeV)/\Delta m^2_5/\sin
2\theta$, they found that the main assumption of \cite{Hiroshi}
about the $\delta$-correlated form of matter density fluctuations,
$L_0\ll l_{osc}$, is violated.  Really, the opposite inequality holds,
$L_0\gg l_{osc}$, especially for typical g-mode wave lengths,
$\lambda_g\sim L_0\gsim 0.1R_{\odot} = 7\times 10^4~km$ {\it if the scale
of random domains is associated with a g-mode wave length}.

Let us turn to the MHD waves trapping in cavities which are appearing in
the central regions of the Sun as a result of simultaneous acting of
magnetic
fields and density inhomogeneities. There are two advantages of these
oscillations:
(i) accumulation of wave energy in a resonance Alfven layer that leads to
a large $\delta \rho/\rho_0$ (see below); (ii) short wave lengths
$\lambda_{MHD}\ll l_{osc}$ that allows us to consider these waves as a
source of $\delta$-correlated matter noise.

First, notice that strictly speaking these MHD waves are called
magneto-acoustic-gravity waves (MAG) \cite{Zhugzhda} (see old
bibliography therein).
The spectrum of MAG waves is very wide and near the photosphere some of
them
can be trapped in the sub-photospheric waveguide with consequent
transformation into the magnetosonic waves characterized by 3-5-min periods
which indeed are observed in the
active regions of the solar atmosphere.

We are interesting here in low-frequency branches of MAG waves which
can lead to long periods (like days or weeks) for temporal variations of
neutrino fluxes that, in general, can be observed in the SuperKamiokande
(SK) experiment.

In the recent paper by Guzzo et al\cite{Guzzo} it was found that in the
presence of gravity force ideal magnetohydrodynamics (MHD) has a
solution for magnetic field perturbation
  with the periods of a few
days that correspond to slow and Alfven waves,
$\omega_S\sim\omega_A= kB_0/\sqrt{4\pi \rho (r)}$ where $k$ is the wave
number (free parameter obeying $k> 2\pi/R_{\odot}$, $R_{\odot}$ is the
solar radius), $B_0$ is a regular magnetic field in the Sun and
$\rho(r)$ is the SSM matter density.

A search for such periodicity in SK could be, in principle, realized.
However, the authors of\cite{Guzzo} confined themselves with the
statement that ``...time fluctuations of the
solar neutrino flux have to be observed if the resonant neutrino
spin-flavor precession solution to the solar neutrino problem works''.

In contrast to \cite{Guzzo}, we show that for the same MAG waves while one
neglects neutrino magnetic moment or RSFP
\cite{RSFP} is absent the MSW solution to SNP is nevertheless
crucially modified and can lead to the corresponding temporal
variations of neutrino flux.

In addition, the authors \cite{Guzzo} did not calculate in a
self-consistent way the location $r_s$ and the width $\Delta r_s$ of
the singular (resonant) layer since they did not solve the eigen
value problem on the base of solutions of the Heun equation (Eq.
(\ref{master}) below) or they did not find the corresponding MAG wave
spectrum (including the imaginary part of frequency connected with
$\Delta r_s$).

We show that low frequences of eigen oscillations correspond to the
resonant layer position deep in the solar interior (under the bottom of
the convective zone) or a cavity for MAG waves being bounded by: (i) the
center of the Sun and (ii) the Alfven resonant layer is localized in the
central region where matter density oscillations can influence the MSW
resonance for neutrinos.

Note that large density fluctuations $\delta \rho/\rho_0$ in these
narrow layers,  $\Delta z_s\ll l_{osc}$, can
influence the MSW conversions like a matter white
noise\cite{Hiroshi} if these layers are tightly placed along neutrino
trajectory.

This paper is organized as follows.

In section 2 we present the total set of MAG equations
and derive the master Heun equation.

In section 3 we find  asymptotic
solutions of master equation near the center (subsection 3.1), at the
surface of the Sun (subsection 3.2) and around the singular Alfven layer
(subsection 3.3).

In subsection 3.4.1 we discuss the cavity model in ideal MHD and
collisionless damping of MAG waves in an Alfven resonant layer.
Then in subsections 3.4.2, 3.4.3 using reasonable
boundary conditions from the common solution in whole region $0\leq r \leq
R_{\odot}$ we derive the dispersion relation for the discrete frequency
 spectrum $\omega_n (B_0, k_x, k_y)$. Here the order of modes $n = 1,2,..$
enumerates the decomposition over the set of MAG eigen waves replacing Eq.
(\ref{simple}) to \begin{equation} \rho (z, t) = \rho_0(z)[ 1 +
Re~\sum_n\delta \rho_n(z, t)/\rho_0(z)]~.  \label{main} \end{equation}

Then in section 4 we calculate the resonance layer location $z_s$ and
its width $\Delta z_s(n)$ making use of the wave energy flux bounded
through
imaginary part of eigen frequency.

In section 5 we present main results for matter density fluctuations that
should be substituted into Eq. (\ref{main}) and calculate the enhancement
of
a density perturbation $\delta \rho_n/\rho_0$ in a resonant Alfven layer.

In final section 6 we discuss the validity of our results and state some
problems for future exploration.

\vskip 0.3cm
\section{Master equation for MAG waves}
\vskip 0.3cm
For MAG waves we use the full set of MHD equations including
the gravity force $\rho \vec{g}$ where $\vec{g} = \nabla \Phi$ and the
gravity potential $\Phi$ is given by the Poisson equation
$\Delta \phi = - 4\pi G\rho$, $G\sim M_{Pl}^{-2}$ is the Newton constant.

The Euler equation (conservation of momentum) is of the form
\begin{equation}
\rho \frac{d\vec{v}}{dt} = - \nabla p  + \rho \vec{g} +
\frac{1}{4\pi }\Bigl [rot~\vec{B}\times \vec{B}\Bigr ]~, \label{Euler}
\end{equation}
where $d/dt = \partial/\partial t + \vec{v}\cdot \nabla$ is the
substantial derivative and the last (Lorentz force) term
 may be transformed to the sum of
the magnetic field pressure term $-\nabla B^2/8\pi$ and a
perturbation magnetic field contribution $-(4\pi )^{-1}\Bigl
([\vec{b}\times [\nabla \times \vec{B}]] + [\vec{B}\times [\nabla\times
\vec{b}]]\Bigr )$ when in linear MHD we decompose the total field $\vec{B}
= \vec{B}_0 + \vec{b}(t,\vec{r})$ (see below subsection 2.1).

 The mass conservation leads to the continuity equation
\begin{equation}
\frac{d\rho}{dt} + \rho u = 0~.
\label{mass}
\end{equation}
Here for compressible gas the function $u = div~\vec{v}\neq 0$  is the
main object for search of solutions with MAG waves.
We decompose density in linear MHD
as $\rho = \rho_0 + \delta \rho(t, \vec{r})$ with the background density
$\rho_0$ given by SSM \cite{Bahcall}.

The energy conservation obeys
\begin{equation}
\frac{dp}{dt} - \gamma\frac{p}{\rho}\frac{d\rho}{dt} = -
(\gamma - 1)Q~, \label{pressure} \end{equation} where the factor $\gamma
= c_p/c_V$ is given by the heat capacities, $Q$ is the rate of all
energy density sources and losses.

Finally system is completed by the Faradey equation for the magnetic field,
$$
\frac{\partial \vec{B}}{\partial t} = rot~[\vec{v}\times \vec{B}] +
\frac{c^2}{4\pi \sigma}\Delta \vec{B}~,
$$
that transits in ideal MHD (the conductivity $\sigma\to \infty$) to the
equation for {\it frozen-in} magnetic field,
\begin{equation}
\frac{\partial \vec{b}}{\partial t} = \nabla \times \Bigl [\vec{v}\times
\vec{B}_0\Bigr ]~.
\label{frozen}
\end{equation}
In the last equation we assumed that the background field $B_0$ does not
depend on time.

Obviously, these equations transit to the Standard Helioseismology
Model (SHM ) equations in the limit $B\to 0$.

The system of linearized Eqs.
(\ref{Euler}-\ref{frozen}) for non-adiabatic perturbations in a
compressible
ideal conducting medium with arbitrary direction of the magnetic field
{\bf B} and in the Cowling approximation (neglecting the
perturbation of the gravitational potential, $\delta \phi = 0$)  is
reduced to
\begin{eqnarray} &&\frac{\partial \delta
\rho}{\partial t} + (\vec{v}\nabla )\rho_0 + \rho_0u= 0~, \nonumber\\
&&\rho_0\frac{\partial \vec{v}}{\partial t} +  \nabla \delta p -
\vec{g}\delta \rho +  \frac{1}{4\pi}\Bigl [\nabla
(\vec{B}_0\vec{b}) - (\vec{B}_0\nabla )\vec{b}\Bigr ] =
0~,\nonumber\\ &&\rho_0c_V\Bigl [\frac{\partial \delta T}{\partial t} +
(\vec{v}\nabla )T_0\Bigr ] + p_0u = \delta Q~,\nonumber\\
&&\frac{\partial \vec{b}}{\partial t} - (\vec{B}_0\nabla )\vec{v} +
\vec{B}_0 u = 0~, \nonumber\\
&&\frac{\delta p}{p_0} = \frac{\delta \rho}{\rho_0} + \frac{\delta
T}{T_0}~.
\label{linear}
\end{eqnarray}
Here $\vec{v}$, $\delta T$, $\delta \rho$, $\delta p$ and $\vec{b}$ are
small Euler perturbations of the velocity, the temperature, the density,
the pressure and the magnetic field strength correspondingly. One
supposes that at initial state the gas is immovable ($\vec{v}_0 = 0$)
and for the constant magnetic field $B_0$
the ideal gas ($p_0 = \tilde{R}\rho_0 T_0$) obeys the hydrostatic
equilibrium, $\nabla p_0 = \rho_0g$. Notice that fifth equation in
(\ref{linear}) follows from equation of state of ideal gas.

In what follows we consider the wave dynamics in central regions of the
Sun where the adiabatic condition, $\delta Q=0$, is fulfilled with good
accuracy for the high frequency oscillations in SHM and MHD. To simplify
the further analysis we adopt this condition for whole spectrum of
oscillations in MHD.

Since we assume the background magnetic field $\vec{B}_0$ has the dipole
structure one expects that it occurs both constant and
uniform near the dipole axis or we can put $B_0 = const$
within central region of the
Sun where a cavity for eigen MAG oscillations is
localized. In addition to, since neutrinos propagate to the Earth along
the ecliptic one can suppose that the dipole magnetic field is
perpendicular to the direction of neutrino momentum.

Thus, we choose the Cartesian coordinate system with z-axis
directed upon opposite direction to the neutrino momentum or along
$\vec{g}$.  This means: (i) $z=0$ at the surface of the Sun and (ii)
$z = R_{\odot}$ at its center. In this system the magnetic
field and acceleration vectors are $\vec{B}_0 = (B_0,0,0)$, $\vec{g} =
(0,0,g(z))$.

In order to obtain an analytical solution of the system Eq.
(\ref{linear}) we adopt additional assumptions based on an effective short
wave length along z-axis, $\lambda_z\ll H_{\rho} = \mid
(1/\rho_0)d\rho_0/dz\mid^{-1}$
where $H_{\rho}\equiv H = 0.1R_{\odot}= const$ is the density height
scale\cite{Bahcall}. Really in most interesting cases of cavities
in deep solar interior
the MHD wave
spectrum $\omega_n$ depends on large order (node) numbers $n\gg 10$ with
the effective wave length $\lambda_z\sim L/n$ where $L\simeq z_s\leq
R_{\odot}$ is the cavity width, $z_s$ is the Alfven resonant layer
position (see below section 4).

On the other hand, using the real model of the Sun \cite{Stix}
one can easily show that both the temperature $T_0(z)$ and the
acceleration $g(z)$ have large height scales, $H_T = \mid
(1/T_0)dT_0(z)/dz\mid^{-1}$ and $H_g = \mid (1/g)d g(z)/dz\mid^{-1}$
correspondingly, obeying the inequalities $H_g,~H_T> H_{\rho}$. As result
we may consider the gravity force acceleration g and the temperature $T_0$
as some constants on scales of MHD wave lengths, $g \approx const$,
$T_0\approx const$. This approximation remains valid even on whole cavity
width $L$ for "narrow" cavities $3H\lsim L\lsim H_T,~H_g$ including the MSW
resonant point $r_{MSW}\sim 0.3R_{\odot}$ in the solar interior while it
fails near the center of the Sun, $z\ll H_{\rho}$, where $g\sim z$.

The isothermal approximation with $g\approx const$ means that the
density and the pressure are given by the exponential (Boltzman) law:
$\rho_0\sim e^{z/H}$, $p_0\sim e^{z/H}$ obeying the ideal gas equation
state $p_0/\rho_0 = RT_0/\mu$ with the gas constant $R = \kappa/m_u$ and
the molecular weight $\mu$ measured in mass units $m_u\simeq m_p$. Here
$\kappa$ is the Boltzman constant; $H = c^2/g = const$ is the height
scale\cite{Bahcall}; $c = \sqrt{RT_0/\gamma \mu} = const$ is the
isothermal sound velocity and the adiabatic sound velocity is given by
$c_s^2 = \gamma c^2 = const$.

Thus we are looking for solutions in the form $\vec{v} =
\vec{v}(z)\exp \Bigl (-i(\omega t - k_xx -k_yy)\Bigr )$
(and analogously for $\delta \rho$ and $\vec{b}$).
We keep here notations of the paper \cite{Dzhalilov2}.
In these notations we obtain from the system (\ref{linear}) all
perturbation functions. In particular, velocity perturbations are of the
form
\begin{equation} v_x = - \frac{ik_x(c_s^2u + gv_z)}{\omega^2}~,
\label{vx} \end{equation} \begin{equation} v_y =
\frac{-ik_y}{\omega^2}\Bigl [gv_z + \Bigl (c_s^2 +
\frac{v_A^2\omega^2}{\omega^2 - k_x^2v_A^2}\Bigr )u\Bigr ]~, \label{vy}
\end{equation}

\begin{eqnarray} v_z = &&\Bigl
(\frac{\omega^2g}{k_{\perp}^2g^2 - \omega^4}\Bigr )\times \nonumber\\
&&\times \{ \Bigl [1 - \frac{k_{\perp}^2c_s^2}{\omega^2} +
\frac{\omega^2(\gamma - 1) - k_y^2v_A^2}{\omega^2 - k_x^2v_A^2}\Bigr ]u
+\nonumber\\ && + \Bigl (c_s^2 +\frac{\omega^2v_A^2}{\omega^2 -
k_x^2v_A^2}\Bigr )\frac{1}{g}\frac{du}{dz}\}~.  \label{vz}
\end{eqnarray}

>From the continuity equation Eq. (\ref{mass}) or from the first
equation in (\ref{linear}) we have obtained the density perturbation,
\begin{equation}
\frac{\delta \rho}{\rho_0} =  \frac{1}{i\omega}\Bigl [ u +
\frac{v_z}{H}\Bigr ]~, \label{contin} \end{equation} and from
Maxwell equations for ideal plasma Eq. (\ref{frozen}) or from the
fourth equation in (\ref{linear}) we find magnetic field perturbations,
\begin{equation} b_x =  - \frac{k_xv_x}{\omega}B_0 +
\frac{u}{i\omega}B_0~, \label{bx} \end{equation} \begin{equation} b_y =
- \frac{k_xv_y}{\omega}B_0~, \label{by} \end{equation} \begin{equation} b_z
= - \frac{k_xv_z}{\omega}B_0~.  \label{bz} \end{equation}

One can easily see from Eqs. (\ref{vx})-(\ref{bz}) that all perturbation
functions are expressed via one unknown function $u(z) = dv_z/dz + ik_xv_x
+ik_yv_y$.

The master equation for $u(z)$ follows from last definition and coincides
\footnote{ More than hundred years this equation for an arbitrary argument
$\xi$ is known in mathematics as the Heun equation \cite{Heun}.}
with  Eq. (3) in \cite{Dzhalilov2}
\begin{eqnarray} &&(\xi - 1)(\xi - a)\xi\frac{d}{d\xi}\xi\frac{du}{d\xi} +
\xi^2(\xi - 2 +a)\frac{du}{d\xi} + \nonumber\\ &&+ \{\xi\Bigl [(\xi - 1)b
+ K_{\perp}^2\Bigl (a + \frac{1}{\gamma K_x^2}\Bigr ) - 1\Bigr ] -
K_{\perp}^2a\}u = 0~, \label{master} \end{eqnarray} where $\xi =
\omega^2/k_x^2v_A^2$ is the independent variable; $v_A(z) =B_0/\sqrt{4\pi
\rho_0(z)}$ is the Alfven velocity; $K_{\perp} = \sqrt{K_x^2 + K_y^2}$ and
$\omega$ are the transversal wave number and the oscillation frequency
correspondingly.

The dimensionless parameters in Eq. (\ref{master}) are given by
\begin{eqnarray}
&&a = 1 - \frac{\Omega^2}{\gamma K_x^2},~~~ b = \frac{\Omega^2}{\gamma}
+ K_{\perp}^2\Bigl ( \frac{\gamma -1}{\gamma\Omega^2} - 1\Bigr
)~,\nonumber\\
&&\Omega = \frac{\omega H}{c},~ K_x = k_xH,~K_y = k_yH,\nonumber\\
&&\xi = \frac{\Omega^2}{\gamma K_x^2}\beta (z);~~\beta (z) =
\frac{c_s^2}{v_A^2} =  \beta_0e^{z/H}~.
\label{parameters} \end{eqnarray}
Eq. (\ref{master}) is the ordinary differential equation of the second
order  with four singular points: $\xi = 0,~1,~a,~\infty$. The
solutions of this equations are given by the Heun
functions\cite{Heun}. These solutions were explored in the
paper\cite{Dzhalilov2}, however, they are too complicated for
application to solar neutrino physics.

Therefore, we have simplified Eq. (\ref{master}) in order to find
solutions through elementary functions and to study main physical
properties of the MAG wave cavity within solar interior.

In general,  when both $K_x\neq 0$ and $K_y\neq 0$, the master
equation Eq. (\ref{master}) has two singular levels:
the resonance level $\xi =a$ arises for the condition
$\omega^2 = c_s^2v_A^2k_x^2/(c_s^2 + v_A^2)$ ({\it casp resonance}) and the
condition $\xi =1$ is fulfilled for $\omega^2 = k_x^2v_A^2$ ({\em
Alfven resonance}).

As shown in \cite{Dzhalilov2} for the solar conditions the casp resonance
level lies in medium above the Alfven level, however, the distance between
these levels, $\Delta z_{Acasp} = H\ln [1 - \Omega^2/\gamma K_x^2]^{-1}$
is negligible for low frequencies $\Omega\ll \gamma K_x^2$ because we are
interesting in 1-10 days variations of density fluctuations with the
probable influence neutrino flux.

When waves reach resonance level which has
a finite width in z-direction and is infinite in both perpendicular
directions they can be efficiently absorbed within resonant layer. This
leads to concentration of wave energy along magnetic field. The energy
absorption coefficient is of the form \footnote{In this formula one
neglects the temporal decay of a wave amplitude since free oscillations
without lower reflector and thereby with zero imaginary part of
the frequency $Im~\Omega = 0$ are assumed.} \cite{Zhugzhda}
\begin{equation} D = 1 - \exp ( -2\pi\sqrt{4b - 1})~, \label{absorb}
\end{equation} or absorption becomes more efficient for low frequencies,
$\Omega\to 0$, $b\to \infty$, $D\to 1$.

The coincidence of casp resonant layers with the Alfvenic ones for low
frequences ($\Omega\ll \gamma K_x^2$)  means that the resonant layer $\xi
=1$ may appear only within the Sun where the condition $v_A\ll c_s$ is
fulfilled that corresponds to the inequality for gas and magnetic
field pressures, $p_{gas} = p_0\gg p_{mag} = B_0^2/8\pi$.

In such conditions we can
simplify our master equation Eq.  (\ref{master}) that can be written in
the final form
\begin{eqnarray}
&&(\xi - 1)^2 \xi\frac{d}{d\xi}\xi\frac{du}{d\xi} + \xi^2(\xi -1)
\frac{du}{d\xi} + \nonumber\\
&&+ \{\xi\Bigl [(\xi - 1)b + K_{\perp}^2\Bigl (1 + \frac{1}{\gamma
K_x^2}\Bigr ) - 1\Bigr ] - K_{\perp}^2\}u = 0~.
\label{master1}
\end{eqnarray}
Here one supposes also $\gamma\approx const$.
\section{Asymptotic solutions }

The main idea how to find simple asymptotic solutions of Eq.
(\ref{master1}) comes from its physical simplifications in three
important regions:  near the photosphere $z = 0$, $\mid \xi\mid\ll 1$,
around the singular (Alfven) layer $\mid \xi - 1\mid < 1$ and near
the center of the Sun, $z = R_{\odot}$ where $\mid\xi \mid\gg 1$  for the
same fixed frequency $\omega$. In these regions solutions are expressed
through elementary functions. Using conditions of linear dependence of
these regional solutions we build  a more general solution that is valid
in whole region $-\infty< z< \infty$.

Then using boundary conditions at the center of the Sun and at the
solar surface we have found eigen frequencies (subsection 3.4).

\subsection{Solution near center of the Sun}
\vskip 0.3cm
In central region of the Sun for $\mid \xi\mid >1$
including the center, $z_s\leq z\leq R_{\odot}$,
where the resonant
Alfven layer position $z_s$ is given by the value
$(\omega/k_xv_A(z_s))^2 = \xi = 1$  we can reduce
Eq. (\ref{master1}) to the
more simpler form
\begin{equation}
\xi^2\frac{d^2u}{d\xi^2} + 2\xi \frac{du}{d\xi} + bu =0~.
\label{inner}
\end{equation}
General solution of this equation is of the form
\begin{equation}
u = u_1 = \frac{1}{\sqrt{\xi}}\Bigl (A_1\xi^{i\nu} + A_2\xi^{-i\nu}\Bigr
)~,
\label{inner1}
\end{equation}
where $\nu = \sqrt{b - 1/4}\approx (K_{\perp}/\Omega)
\sqrt{(\gamma - 1)/\gamma}$, $A_{1,2}$ are arbitrary constants
which are determined from boundary conditions.
For low frequencies $\mid \nu \mid\gg 1$ the derivative is substituted by
\begin{equation}
\frac{du_1}{d\xi} \approx \frac{i\nu}{\xi\sqrt{\xi}}\Bigl (A_1\xi^{i\nu} -
A_2\xi^{-i\nu}\Bigr )~.
\label{derinner}
\end{equation}
Since the variable $\xi$ is proportional to the SSM density profile
$\xi\sim e^{z/H}$ \cite{Bahcall} the solution Eq. (\ref{inner1})
describes the incident wave and the reflected one with the
wave number $k_z = \nu/H$.

One can easily see that the asymptotic Eq. (\ref{inner}) is similar with
the SHM equation for g-modes where the wave number $b/H^2\approx
k_x^2g(\gamma - 1)/\gamma H\omega^2$ in Eq. (\ref{parameters}) includes
the buoyancy frequency term $N^2/\omega^2\sim g(\gamma - 1)/H\omega^2$
coming in SHM equations. This corresponds to the limit $B_0\to 0$ when the
variable $\xi$ increases ($\xi\to \infty$).

\vskip 0.3cm

\subsection{Solution in vicinity of a resonant Alfven layer}

For $\xi\to 1$ Eq. (\ref{master1}) transits to
\begin{eqnarray}
&&(\xi - 1)^2 \frac{d^2u}{d\xi^2} + (\xi -1)\frac{du}{d\xi} + \nonumber\\
&&+ \Bigl [(\xi - 1)\frac{(\gamma -
1)}{\gamma}\frac{K_{\perp}^2}{\Omega^2} + \frac{K_{\perp}^2}{\gamma
K_x^2} - 1\Bigr ]u = 0~.
\label{singular}
\end{eqnarray}
Two linear independent solutions of this equation are expressed through the
Bessel functions of the 1-st kind:
\begin{equation}
u = u_2 = C_1J_{\mu}(\Psi ) + C_2J_{-\mu}(\Psi )~,
\label{singular1}
\end{equation}
where we denoted the argument and the index of Bessel functions
as $\Psi = 2\nu\sqrt{\xi - 1}$,
$\mu = 2[1 - K_{\perp}^2/\gamma K_{x}^2]^{1/2}$
correspondingly and $C_{1,2}$ are arbitrary constants.

For the case of MAG wave propagation along magnetic field ($k_y = 0$)
the index $\mu$ is real, $\mu^2 = 4(\gamma - 1)/\gamma$, while for the
particular case of oblique wave propagation $k_y/k_x =
\sqrt{15\gamma/16 -1} = 3/4$ (ideal H-atomic gas, $\gamma = 5/3$)
the real index
$\mu = 1/2$ corresponds to the trigonometrical functions instead of
the Bessel ones,
\begin{equation}
u_2 = \sqrt{\frac{2}{\pi \Psi}}\Bigl
[C_1\sin \Psi + C_2\cos \Psi\Bigr ]~.
\label{trig}
\end{equation}
As it follows from the decomposition of Bessel functions near the
resonance level $\xi\approx 1$, or explicitly for the particular case
(\ref{trig}) the second independent solution diverges in the
limit $\xi\to 1$.  However, since this singular behaviour is an
intrinsic property of the solution Eq. (\ref{singular1}) caused by
physical reasons we may not exclude this singularity by the forced
condition $C_2 =0$.  It is important for our problem to have
convergence (regular solution) in all
points along integration path $\mid z\mid\to \infty$. An analytic
continuation (transition to the complex argument $\xi = \xi_1 + i\xi_2$)
originated by a physical mechanism discussed
below means that $\xi =1$ is the removable singularity.

Notice that the solution Eq.(\ref{singular1}) is oscillating along
the direction to the center, $\xi\gsim 1$,  and has exponential
behaviour in opposite direction (to the surface of the Sun), $\xi\lsim 1$.
First derivative of the solution Eq. (\ref{singular1})  is of the
form
\begin{equation}
\frac{du_2}{d\xi} =
\frac{2\nu^2}{\Psi }\Bigl [C_1J^{'}_{\mu}(\Psi ) + C_2J^{'}_{-\mu}(\Psi
)\Bigr ]~,
\label{dersingular}
\end{equation}
where $J^{'}_{\mu}(\Psi)
=(\mu/\Psi)J_{\mu}(\Psi) - J_{\mu + 1}(\Psi)$.
\vskip 0.3cm
\subsection{Solution near surface of the Sun}
\vskip 0.3cm

The argument $\xi = \omega^2/(k_x^2v_A^2)$ in Eq. (\ref{parameters}) is
given by the parameter $\beta = \beta_0e^{z/H}$ where
near the solar surface $z = 0$ the factor $\beta_0$ is changing in a
wide region\cite{Priest2}
\begin{eqnarray}
    \beta_0 =\Bigl \{     \matrix{40,&for&B= 1000~G,\nonumber\\
              400,&for&B= 100~G,\nonumber\\
           40000,&for&B= 10~G.}
\label{beta0}
\end{eqnarray}

However, for any $\beta_0$ and $K_x\geq 1$ the argument $\xi$ occurs
small near photosphere, $\mid \xi\mid\ll 1$, since we have considered
low frequencies through whole solar interior, $\omega\lsim
10^{-5}~s^{-1}$. As result the master equation Eq. (\ref{master1}) transits
near photosphere to
\begin{equation}
\xi^2\frac{d^2u}{d\xi^2} + \xi\frac{du}{d\xi} - \Bigl
(\xi\frac{\gamma -1}{\gamma}\frac{K_{\perp}^2}{\Omega^2} +
K_{\perp}^2\Bigr )u = 0~,
\label{surface}
\end{equation}
which has the solution
\begin{equation}
u = u_3  = D_1J_{2K_{\perp}}(y) + D_2J_{-2K_{\perp}}(y)~.
\label{surface1}
\end{equation}
Here the Bessel functions depend on the argument $y = 2\nu\sqrt{-\xi}$,
$D_{1,2}$ are new integration constants. The demand of regularity
of the solution Eq. (\ref{surface1}) on the photosphere, $\xi\to 0$,
results in $D_2 = 0$. The solution Eq. (\ref{surface1}) describes
evanescent oscillations with an exponential decreasing amplitude at the
surface of the Sun.

The derivative of the solution Eq. (\ref{surface1}) is given by
\begin{equation}
\frac{du_3}{d\xi} = - D_1\frac{2\nu^2}{y}\Bigl
[\frac{2K_{\perp}}{y}J_{2K_{\perp}}(y) - J_{1 + 2K_{\perp}}(y)\Bigr ]~.
\label{dersurface}
\end{equation}
\vskip 0.3cm
\subsection{Spectrum of MAG waves}
\vskip 0.3cm
\subsubsection{Cavity model in ideal MHD}
\vskip 0.3cm
We have shown that the Alfven resonant level $\xi =1$ divides whole wave
propagation space into two parts in which MHD waves have distinct
properties.  In the region $1<\xi< +\infty$ the solutions $u_1(\xi)$ and
$u_2(\xi)$ describe waves that propagate in both directions along z-axis
while in the region $0\leq \xi<1$ the solutions $u_2(\xi)$ and $u_3(\xi)$
describe evanescent oscillations that decrease in an exponential way to
the surface of the Sun, $z=0$. The level $\xi = 1$ itself is the singular
level where $u_2(\xi )\to \infty$.

Therefore, waves which are propagating from the center of the Sun reach
the resonance level $\xi = 1$ where their partition implies reflection
back and partial capture at the resonance layer. In general, high wave
amplitudes at the resonance imply a start-up of different dissipation
mechanisms (linear or non-linear) within that layer.
The condition
$v_z = 0$ for the hard center of the Sun is a
natural demand due to high pressure there and leads to full wave
reflection from the center resulting in a cavity
appearance with the lower boundary at the center $z = R_{\odot}$ and
the upper boundary $z= z_s$ at the Alfven resonance layer.

The cavity is defined within the region $1\leq \xi\leq \xi_{\infty}$
and its width $L = H\ln \mid \xi_{\infty}\mid$ is given by the argument
value at the center of the Sun, $\xi_{\infty} =
(\Omega^2/\gamma K_{x}^2)\beta_0e^{R_{\odot}/H}$. Notice that the width L
depends significantly on the value of a central magnetic field via the
parameter $\beta_0$ in Eq. (\ref{beta0}).

The waves which are trapped within cavity loose wave energy due
to partial absorption of their energy in the resonance Alfven layer
with the rate given by an imaginary part of the eigen frequency $Im~(\omega
)\neq 0$.

This dissipationless (i.e. reversible) damping is
analogous to the collisionless mechanism of coherence losses for eigen
(Alfvenic) modes considered in \cite{Ionson} for the Alfven resonant
layer which is the sheath of a solar coronal loop.
The resonant absorption of Alfvenic
surface waves leads to the heating of solar coronal loops.
Moreover, the inertia of electrons $m_e\neq \infty$
($v_{eT}=\sqrt{T/m_e}\neq 0$) should be taken into account or kinetic
approach for wave-particle interaction is necessary to explain
collisionless damping.

The same collisionless damping of MHD waves with absorption of waves
in the Alfven singular layer as a mechanism of heating of laboratory
plasma was considered in \cite{Hasegawa} and in \cite{Priest} - for
heating of a nonuniform cosmic plasma.

In ideal MHD for a
non-uniform medium one can also obtain the collisionless damping of eigen
modes from the dispersion relation solving the corresponding eigen value
problem with given boundary conditions\cite{Ionson}.

There is another analogy that is well-known from plasma physics where the
collisionless decrement of plasma waves comes both from
dispersion relation for eigen modes and from the kinetic theory
beyond the ideal MHD.  The collisionless damping and increment are equal
to each other for rates of the direct (inverse) $\check{C}$erenkov
emission (absorption) $e\leftrightarrow e\gamma$ in the wave-particle
interaction.

The Landau damping of the Alfven waves $\omega = k_xv_A(z_s)$ in
the inverse $\check{C}$erenkov process , $\gamma_Ae\to e$, starts only
within the Alfven resonant layer $\xi =1$ obeying the standard
$\check{C}$erenkov condition $\omega = \vec{k}\cdot \vec{v}_e$ where
the electron velocity $v_e$ is much bigger than the Alfven one, $v_e\sim
v_{eT}\gg v_A$.  This means that those electrons which are moving almost
across Alfven layer (in ecliptic) could absorb Alfven waves .

Thus, accounting for trapping of waves in the cavity $1\leq
\xi\leq \xi_{\infty}$ with partial losses of the wave energy within its
upper boundary ($\xi = 1$) one can solve the problem of infinite wave
amplitude at the resonance singular layer in framework of ideal MHD
without taking into account of usual dissipation mechanisms (viscosity
through collisions, ohmic dissipation,etc..).  Really, for complex
frequency the argument $\xi$ becomes complex or the singularity $\xi =1$
is removable (see below).

Our main task in present
work is the derivation of {\it complex eigen frequencies and eigen
functions} in a cavity we have considered. First step for that is the
construction of a common solution in the whole variable region $\mid
\xi\mid<\infty$.

\vskip 0.3cm
\subsubsection{Common solution and continuity conditions}
\vskip 0.3cm

Let us assume that the complex frequency $\Omega = \Omega_1 +
i\Omega_2\equiv \Omega_1(1 + id)$ obeys the stability condition
$\Omega_2\leq 0$ ($\Omega_1^2>\Omega_2^2$) and input the following
notations
for the independent variable $\xi = \xi_1 + i\xi_2 = \xi_1(1 +
i\zeta)$ :
\begin{eqnarray}
&&\xi_1 = \frac{\Omega_1^2 - \Omega_2^2}{\gamma K_x^2}\beta (z) =
\frac{\Omega_1^2}{\gamma K_x^2}(1 - d^2)\beta (z)\nonumber\\
&&\zeta = \frac{\xi_2}{\xi_1} = \frac{2\Omega_1\Omega_2}{\Omega_1^2 -
\Omega_2^2} = \frac{2d}{1 - d^2}~.
\label{notations}
\end{eqnarray}
Notice that transition to free oscillations corresponds to the limit $d
= \Omega_2/\Omega_1\to 0$.

In the Fig. 1 we plot definition region for all three solutions upon
the axis $\xi_1 = Re~\xi$ where the region $0\leq \xi_1\leq
\xi_{1\infty}$ is the whole integration region. In common regions for
overlap of solutions $u_{1,2,3}(\xi)$ we can use analytic continuation
and connect these solutions each other.

By continuity of functions $u_{1,2,3}$ and of their derivatives
$du_{1,2,3}/d\xi$ we choose two matching points in two overlap
regions : (i) $\xi_1 = 1/2$ for the pair of solutions $u_3(\xi)$ and
$u_2(\xi)$ and (ii) $\xi_1 = 3/2$ for the second pair $u_1(\xi)$ and
$u_2(\xi)$.  Then at these points we obtain  the continuity conditions
\begin{eqnarray} &&u_3(\xi ) = u_2(\xi );~~~~~\frac{du_3(\xi )}{d\xi} =
\frac{du_2(\xi )}{d\xi}~~~~~for~~ \xi_1 = \frac{1}{2}~,\nonumber\\
&&u_1(\xi ) = u_2(\xi );~~~~~\frac{du_1(\xi )}{d\xi} = \frac{du_2(\xi
)}{d\xi}~~~~~for~~ \xi_1 = \frac{3}{2}~.
\label{connection}
\end{eqnarray}
For the complex matching points chosen above,
\begin{equation}
\xi_0 = \frac{1}{2}(1 + i\zeta),~~~~~\xi_*= \frac{3}{2}(1 +
i\zeta)~,
\label{common}
\end{equation}
the arguments of functions in Eqs. (\ref{singular1}), (\ref{surface1})
are of the form
\begin{eqnarray}
&&\Psi_0 = \Psi |_{\xi_1 = 1/2} = 2\nu\sqrt{\xi_0 - 1},~~y_0 =
y|_{\xi_1 = 1/2} = 2\nu\sqrt{- \xi_0},\nonumber\\
&&\Psi_* = \Psi |_{\xi_1 = 3/2} = 2\nu\sqrt{3\xi_0 - 1}~.
\label{common1}
\end{eqnarray}
Substituting the solutions $u_{1,2,3}$ and their derivatives
$u^{'}_{1,2,3}$ given by Eqs.
(\ref{inner1}),~(\ref{derinner}),~(\ref{singular1}),
~(\ref{dersingular}),~(\ref{surface1})~,(\ref{dersurface}) into the
continuity conditions Eq. (\ref{connection}) we obtain finally the
system of algebraic transcendental equations
\begin{eqnarray}
&&C_1J_{\mu}(\Psi_0) + C_2J_{ -\mu}(\Psi_0) =
D_1J_{2K_{\perp}}(y_0)~,\nonumber\\
&&C_1J^{'}_{\mu}(\Psi_0) + C_2J^{'}_{ -\mu}(\Psi_0) =
- D_1\frac{\Psi_0}{y_0}J^{'}_{2K_{\perp}}(y_0)~,\nonumber\\
&&C_1J_{\mu}(\Psi_*) + C_2J_{-\mu}(\Psi_*) =
\frac{1}{\sqrt{\xi_*}}\Bigl (A_1\xi_*^{i\nu} + A_2\xi_*^{-i\nu}
\Bigr )~,\nonumber\\
&&C_1J^{'}_{\mu}(\Psi_*) + C_2J^{'}_{-\mu}(\Psi_*) =
\frac{i\Psi_*}{2\nu\xi_*\sqrt{\xi_*}}\Bigl (A_1\xi_*^{i\nu} -
A_2\xi_*^{-i\nu} \Bigr )~,
\label{algebra}
\end{eqnarray}
from which we can define all arbitrary constants except of one, for
instance, we choose $D_1\neq 0$. Let us recall that
in the solution near surface of the Sun in Eq. (\ref{surface1}) we used
 the boundary condition $D_2 =0$ excluding irregular part of solution.

Notice that in low frequency approximation $\Omega\ll 1$ the index $\nu$
is large, $\mid \nu\mid\gg 1$ (see Eq. (\ref{inner1})), resulting in
large arguments of Bessel functions: $\mid \Psi_*\mid>\mid
\Psi_0\mid\gg 1$, $\mid y_0\mid\gg 1$. As result we substitute
Bessel functions by their asymptotical expansions for large
arguments.
\vskip 0.3cm
\subsubsection{Dispersion equation and spectrum of MAG waves}
\vskip 0.3cm
Let us turn to the boundary condition at the center of the Sun,
$v_z|_{z = R_{\odot}}=0$. From the Eq. (\ref{vz}) in the limit $\mid
\xi\mid\gg 1$ we obtain the ratio for coefficients $A_{1,2}$ in the
solution Eq. (\ref{inner1}) for central region of the Sun:
\begin{equation}
\frac{A_1}{A_2} = - \xi_{\infty}^{-2i\nu}~.
\label{ratio}
\end{equation}
Thus, using the last ratio and the system Eq. (\ref{algebra}), we
obtain the dispersion equation for the MAG waves in the Sun:
\begin{equation}
\frac{\Psi_*}{2\nu\xi_*}\frac{(\varepsilon - 1)}{(\varepsilon + 1)} = 1~,
\label{dispersion}
\end{equation}
where the complex parameter $\varepsilon = \varepsilon_1 +
i\varepsilon_2 = - (\xi_*/\xi_{\infty})^{2i\nu}$ can be expressed for
small damping decrement ($d^2\ll 1$) via real functions
\begin{equation}
\varepsilon_1 = - e^{d\Phi}\cos \Phi~,~~\varepsilon_2 = - e^{d\Phi}\sin
\Phi~.
\label{auxiliary}
\end{equation}
Here the  parameter $\Phi = \Phi (\Omega_1, k_{\perp}, \alpha , B_0)$
depends on the {\it real part} of the eigen frequency $\Omega_1$:
\begin{equation}
\Phi = 2\nu_*\ln \Bigl (\frac{3\gamma
K_x^2}{2\Omega_1^2\beta_0}e^{-R_{\odot}/H}\Bigr )~,
\label{real}
\end{equation}
where $\nu_* = Re~(\nu) = \sqrt{(\gamma -
1)/\gamma}(K_{\perp}/\Omega_1)$ is the real part of the index $\nu$;
$\tan \alpha = K_y/K_x$ is different from zero for oblique wave
propagation, $K_y\neq 0$.

As result of the parametrization $\varepsilon_{1,2}(d, \Omega_1)$ in
Eq. (\ref{auxiliary}) the dispersion equation Eq.(\ref{dispersion}) can be
rewritten as the system of transcendental equations for the real
($\Omega_1$) and the imaginary ($d$) parts of eigen frequency
\begin{eqnarray} &&d(\varepsilon_1 - 1)  + \varepsilon_2\Bigl (1 -
\frac{3}{\sqrt{2}}\Bigr ) = 0~,\nonumber\\ &&d\varepsilon_2 + 1 -
\varepsilon_1 + \frac{3}{\sqrt{2}}(1 + \varepsilon_1) = 0~.
\label{dispersion1} \end{eqnarray} One can easily see that $d=0$ is not a
solution of Eq.  (\ref{dispersion1}).  This means that full reflection
from resonance layer never happens.  Accounting for notations in Eq.
(\ref{auxiliary}) we obtain from Eq. (\ref{dispersion1}) another form of
dispersion equations \begin{eqnarray} &&\Phi \tan \Phi =
\frac{6\sqrt{2}}{7}\ln \frac{7}{11 - 6\sqrt{2}}~,\nonumber\\ &&d =
\frac{1}{\Phi}\ln \frac{7}{11 - 6\sqrt{2}}~, \label{dispersion2}
\end{eqnarray} where $\Phi$ is given by Eq. (\ref{real}).

Since a solution of the first equation in Eq. (\ref{dispersion2})
depends on the harmonic number $n = 1,2,...$ only, $\Phi = \Phi (n)$ , the
damping decrement $d$ depends on that harmonic number too, $d = d(n)$.
This dependence is plotted in Fig. 2. The calculation gives $d< 0$ for
positive values $n\geq 1$.
%\vskip0.3cm
%\psfig{file=d.ps,height=8cm,width=8cm}\\
%\vskip0.3cm
One can see from Fig. 2 that number of harmonics is
finite: $1\leq n\leq n_{max}$. While $\mid d\mid$ is small for a
large $n$ the nonzero value $d\neq 0$ for $n_{max}$  means that energy
losses in cavity always happen due to resonance absorption.

The phase velocity of waves along the magnetic field $\vec{B}_0$
normalized to the sound velocity $V_{ph}/c_s = \omega_1/k_xc_s =
\Omega_1/\sqrt{\gamma}K_x$ is the second solution of dispersion equations
Eq. (\ref{dispersion2}) as it follows from the definition of the function
$\Phi$ in Eq.  (\ref{real}).  This solution depends on all set of
parameters:  the harmonic number $n$, the value of magnetic field $B_0$,
the wave angle of incidence $\alpha$, $\tan \alpha = k_y/k_x$, the
adiabatic parameter $\gamma$ and the wave number $k_x$. In Fig. 3a we show
the dependence $V_{ph}(n)$ on $n$ for different values of the magnetic
field $B_0$ and for a particular case of the longitudinal wave
propagation, $\vec{k}\parallel \vec{B}_0$, $k_y = 0$.

For a given $n$ the dispersion equation has two solutions: the high
frequency one (plotted by the solid line in Fig. 3a) and the low frequency
one (dashed line). The cross point of these curves determines the maximal
value $n_{max}$. The value $n_{max}$ decreases when the magnetic field
$B_0$ increases. The analogous dependence $\omega_1(n)$ for the fixed
magnetic field value $10~G$ and for different wave propagation angles
($k_{\perp}/k_x = 1,2,5$ corresponds to $\alpha = 0^o,60^o,78^o$) are shown
in Fig. 3b. The node number $n$ increases with increase of $\alpha$.

The change of a node number n in dependence on $B_0$, $\alpha$
is connected with the change of a cavity width for these varying
parameters.  Combining results in Figs. 2,3 we can find the imaginary part
of the eigen frequency $\Omega = \Omega_1 + i\Omega_2$, $\Omega_2(n) =
d\Omega_1$, for a given harmonic n.

The peculiarities of the eigen modes $\omega_1 (n, B_0, k_x, \alpha)$ are
seen from Figs. 3a, 3b: the high frequency branch (solid lines) has
the minimum value $\omega_{min}^{(high)}(n_{max})$  for $n_{max}$ and the
maximum one $\omega_{max}^{(high)}(n = 1))$  for $n = 1$. Vice versa,
the low frequency branch (dashed lines) has the minimal value
$\omega_{min}^{(low)}(n = 1)$ for $n =1$ and the maximal one
 $\omega_{max}^{(low)}(n_{max})$ for $n =n_{max}$. Note that the first
 harmonic ($n = 1$) for the high frequency branch corresponds to the
 magnetosonic waves: $V_{ph}\sim c_s$ that is outside of our approximation
 of low frequences, $\omega\ll k_xc_s$. As result we got too fast damping
 of such waves, $\mid d\mid\gsim 10^{-2}$.

We do not consider here the low frequency branches (dashed lines) for
which the rotation of the Sun becomes important. This rotation changes
crucially properties and spectra of low frequency oscillations ($T\geq
T_{\odot}$) in the Sun as it was found in \cite{Dzhalilov}.

Note that since a node number $1\leq n\leq
n_{max}$ is finite a continuum does not arise, spectrum is always
discrete (ridge structure).

The spectrum $f =\omega_1(n, B_0, k_x, \alpha=0)/2\pi$
(in the dimension units $\mu Hz$) is shown in Fig. 4 for the longitudinal
wave propagation: we plotted the real part of frequencies $\omega_1(n, B_0
= 10~G, k_x)$ in dependence on the wave number $k_xR_{\odot}$ for the
particular case $B_0 = 10~G$, for $k_y = 0$ and for different harmonics
$1\leq n\leq n_{max} = 2780$.  The curve $n = 2780$ is common for low and
high frequency branches. Since the long period $f^{-1} = (1~\mu Hz)^{-1} =
11,6~days$ is at the boundary of our approach we conclude that all
dashed lines in Fig.  4
with $n \leq n_{max} = 2780$ and some part of solid lines ($n \gsim 2000$,
or $f\lsim 1~\mu Hz$) correspond to too low frequencies when we need to
take into account the Sun rotation (with the period $T_{\odot} \simeq
27~days)$.

\section{The position $z_s$ and the width $\Delta
z_s$ of a resonant layer} \vskip 0.3cm

The appearance of a resonant Alfven layer within the
Sun, $0\leq z_s\leq R_{\odot}$ can lead to an interesting application for
SNP. From the condition $\xi_1 = 1$ substituted into Eq. (\ref{notations})
we have found the distance of a singular layer from the center of the Sun
for the case $d\ll 1 $:  \begin{equation} \frac
{z_s}{R_{\odot}} = 1 - \frac{z_{1s}}{R_{\odot}} = 1 - \frac{1}{10}\ln
\Bigl (\frac{\gamma K_x^2}{\beta_0\Omega_1^2}\Bigr ) = 1 -
\frac{1}{10}\ln \Bigl (\frac{c^2_s}{\beta_0v_{ph}^2}\Bigr )~,
\label{distance} \end{equation}
where $z_{1s}$ is the distance of the same singular point measured from
the surface at $z = 0$.  In Fig. 5 (a,b) we show the harmonic number
dependence $z_s(n)$ for given values of the magnetic
field $B_0$ and the wave incidence angle $\alpha$.

One can see that for any values of the parameters $B_0$, $\alpha$ and for
the low frequency branch (dashed curves) the resonant layer is near the
center of the Sun ($z_s/R_{\odot}\leq 0.25$). For the high frequency
branch (solid curve) it occurs at the edge of the solar core relevant
to the MSW resonance region and either in the radiative zone or in the
convective one ($0.25\leq z_s/R_{\odot}\leq 1$). With increase of the
magnetic field $B_0$ the acceptable node number $n\leq n_{max}$ decreases
(see Fig.  5a for fixed $k_y = \alpha = 0$) while for the fixed magnetic
field value $B_0 = 10~G$ it increases if $\alpha$ increases (Fig. 5b).

Let us turn to the derivation of the resonant layer
width $\Delta z_s$ where a significant part of the wave energy  is
concentrated.  For this goal we have calculated the density of the
hydrodynamical energy flux across an Alfven resonant layer
\begin{equation} P^{(mech)}_z (\xi_1(z)) =
\frac{1}{2} Re (\delta p\bar{v}_z^*)) =
\frac{p_0}{2}Re~\Bigl [\frac{v_z^*}{i\omega}\Bigl (\frac{v_z}{H} + \gamma
u\Bigr )\Bigr ] ~,
\label {flux} \end{equation} where
the pressure perturbation $\delta p = (p_0/i\omega )(v_z/H + \gamma u)$ is
obtained from the system Eq. (\ref{linear}), $v_z^*$ is the complex
conjugate value of $v_z$ and the function $u (z) = u_2(z)$ is given by the
Eq.  (\ref{singular1}).

Notice that we have neglected electromagnetic energy flux density of waves
across resonant layer because near that Alfven layer ($\xi \sim 1$  )
the z-component of the Poynting vector is much smaller than the mechanical
one, $P_z^{(em)}\ll P_z^{(mech)}$, since the condition $p_{gas}\gg
p_{mag}$ is fulfilled within the Sun.

In the case of free oscillations the imaginary part of
the eigen frequency is zero, $\Omega_2 = d =0$, and the flux
$P_z^{(mech)}(\xi)$ is infinite at the singular layer, $P_z^{(mech)}(\xi =
1) = \infty$.  However, for eigen modes bounded within cavity ($d\neq 0$)
this flux is finite, $P_z^{(mech)}(\xi_1 =1)\neq \infty$, and the function
$P_z^{(mech)}(\xi)$ has a finite maximum for the argument $\xi_1 = 1$.

The height of this peak is determined by the value of $d =
\Omega_2/\Omega_1$. In order to find the half-width of that peak we have
used the equation $P_z^{(mech)}(\xi_1(z)) = P_z^{(mech)}(\xi_1 =1)/2$
which has two solutions: $\xi_1^-<1$ and $\xi_1^+>1$.

For the half-width we have found \begin{equation} \Delta z_s = H\ln
\frac{\xi_1^+}{\xi_1^-}~.  \label{width} \end{equation}

In Fig. 6 we show
the dependence of the width $\Delta z_s$ on node numbers n for
different values of the background magnetic field $B_0$ and for the
longitudinal wave propagation, $k_y= 0$. One can see that the width
$\Delta z_s$ becomes more narrower for large node numbers lowering to
$\Delta z_s\lsim 1~km$ for the case $B_0 = 10~G$ in the region $n\geq
1000$ (radiative zone, see Fig. 5a where the corresponding position
$z_s(n)$ is shown).

We have found
too narrow spikes (hundreds meters) near the node number $n\sim
2680$ in the most interesting region $z= r_{MSW}\sim 0.3R_{\odot}$ for
the same ambient magnetic field $B_0 = 10~G$ while for the field $B_0 =
100~G$ the width reaches $\Delta z_s\sim 5~km$ for the same position $z_s$
with the node number $n\sim 268$ (compare Figs. 7a, 7b below).

In reality the position $z_s(n)$ is discrete in dependence on the node
number n (it looks continuous in large-scale Figs. 5a, 5b).
For the high frequency branch for which our approximation of
the Sun at rest seems to be valid we have shown in Figs.  7a, 7b  (for the
background magnetic field $B_0 = 10~G$ and $B_0 = 100~G$ correspondingly)
the distance between neighbouring resonant Alfven layers for the most
interesting region near the MSW resonant point ($r_{MSW}\sim
0.3R_{\odot}$).  One can see that the distance between layers $n$ and
$n+1$ is changing in order of magnitude from $z_s(n) - z_s(n +1)\simeq
200~km$ ($B = 10~G$) up to  $z_s(n) - z_s(n +1)\simeq 2000~km$ ($B =
100~G$).

\vskip 0.3cm

\section{Density fluctuations}

\vskip 0.3cm

Considering whole solar interior $0< z< R_{\odot}$ for the same background
magnetic field values $B = 10~G$ and $B = 100~G$ we have found many
resonant layers shown in our main Figures 8a, 8b where an
unnormalized density fluctuation $\delta \rho (z) /\rho$ (calculated from
Eq.  (\ref{contin})) is plotted as the function of z, $0\leq z\leq
R_{\odot}$.

Notice that a height of spikes occurs the random one because of
uncertainty of numerical calculation near each resonant layer when
the numerical code is running along whole solar radius with many spikes
skipping occasionally some sharp maxima.

Obviously from comparison of Figs. 8a, 8b that in the case of the small
magnetic field $B_0 = 10~G$ spikes are placed more tightly than in the
case of $B_0 = 100~G$ and really resemble a density matter noise since
small distance between neighbouring resonant layers shown in Fig. 7a,b.

Let us estimate the
enhancement of the density perturbation $\delta \rho/\rho_0$ given by Eq.
(\ref{contin}), \begin{equation} A = \frac{\delta \rho/\rho_0)_{z =
z_s}}{(\delta \rho/\rho_0)_z} \simeq \frac{\mid v_z(\xi)/H +
u_2(\xi)\mid_{z = z_{1s}}} {\mid v_z(\xi)/H + u_2(\xi)\mid_z} ~,
\label{enhance}
\end{equation} in the vicinity of an Alfven resonant layer, $z -
z_{1s}\ll H$, obeying the inequalities
$$
(\xi_1 - 1)\ll   \frac{v_A^2}{c_s^2} =(\beta_0 e^{z/H})^{-1}\ll 1,
$$
where the real part of the $\xi$-argument equals to $\xi_1 = e^{(z -
z_{1s})/H}$ (here $\xi_1 >1$ since $z >z_{1s}$) and the first inequality
means a finite small displacement from the resonant point $z_{1s}$ provided
the large parameter $\beta_0 = 4\times 10^4\gg 1$ for the small magnetic
field value $B_0 = 10~G$ is substituted.

For simplicity let us also consider the longitudinal wave propagation
$\vec{k}\parallel \vec{B}_0$, $k_y = 0$, when the index $\mu$ is real, or
for hydrogen with $\gamma = 5/3$ we find $\mu = 2\sqrt{(\gamma -
1)/\gamma}\sim 1.27$.

One can easily check from Eq. (\ref{vz}) that for conditions chosen above
the longitudinal speed component is of the order
\begin{equation}
v_z(\xi)/H\approx -\gamma
u_2(\xi) + O(\beta_0^{-1})~,
\label{approx}
\end{equation}
 where the singular solution $u_2(\xi)$ in Eq.
(\ref{singular1}) takes of the asymptotic form $u_2(\xi)\simeq C_2J_{ -
\mu}(\Psi)\approx C_2(\nu \sqrt{\xi - 1})^{-\mu}/\Gamma (\mu +
1)$.

Thus, the amplification Eq.  (\ref{enhance}) of a MAG wave density
amplitude $\delta \rho/\rho_0$  is given by \begin{equation} A\simeq \Bigl
[\frac{((z - z_{1s}(n))/H)^2 + 4d^2(n)}{4d^2(n)}\Bigr ]^{1 + \mu/4}~,
\label{enhance2} \end{equation}
or enhancement is really big because of transparency for MAG
eigen waves shown in Fig.  2, $d(n) = \Omega_2/\Omega_1\ll 1$, especially
for large harmonics $n\gg 1$.

In Figs. 9a,b we have shown the profile $\delta \rho_n(z)/\rho_0$ of the
selected mode n = 300 built numerically in the idealized case $B_0 = 10~G$
and $k_y =0$ for which the resonant layer position occurs just under the
top of the convective zone, $z_s/R_{\odot} = 1 - z_{1s}/R_{\odot}= 0.9985 $
(in our model we neglect turbulent properties of medium in that
region).

One can easily understand the continuous exponential increase of the wave
$n = 300$ propagating from the center $z=R_{\odot}$ up the resonant
Alfven layer $z_{1s}\sim 0.01$ near the photosphere (see Fig. 9b). This
follows from the solution Eq.  (\ref{inner1}) extended on whole
solar interior and rewritten as $u_1(z) = e^{-
z/2H}(A_{10}e^{iz\sqrt{b}/H} + A_{20}e^{-iz\sqrt{b}/H})$.

For this largest
cavity the wave length of MAG waves along z-axis equals to $\lambda_z\sim
z_s/300\sim 2300~km$ and the enhancement seen from Fig.  9a (for
unnormalized amplitude) is of a few orders of magnitude ( $A\sim 10^2-
10^3$ for the distance $\lambda_z\sim (z - z_{1s})$) in good agreement
with the analytic estimate Eq.  (\ref{enhance2}).

>From the right side of that resonant layer one can see in Fig. 9a a steep
decrease of density perturbation described by Eq. (\ref{contin}) through
the solution Eq.  (\ref{surface1}) vanishing exponentially to the surface
$z = 0$.

\vskip 0.3cm

\section{Discussion}
~~~~~We have studied MHD (or more correct name: MAG) wave origin of the
solar matter density fluctuations $\delta \rho/\rho_0$ that could be
important both for Solar Physics and for the MSW solution to SNP.

We have found MAG-wave eigen modes for cavities within solar interior with
the spectrum given by the dispersion equations Eq. (\ref{dispersion2}).
The lower boundary of cavities is the common point - center of the Sun,
$z = R_{\odot}$, where the longitudinal perturbations are absent, $v_z =
b_z = 0$, and the top (upper) boundary is given by the Alfven resonant
layer position, $z= z_s(n)$, that depends on the node number $n$, the
background magnetic field value $B_0$ and the wave numbers $k_x,~k_y$.

A finite amplitude of the density perturbation $\delta \rho/\rho$ in each
Alfven resonant layer is determined by the mechanism like the
collisionless damping of MAG waves that is similar with the Landau damping
of Alfven waves = absorption by electrons at the $\check{C}$erenkov
resonance $\omega = k_xv_A = kv_e\cos \theta$, $v_e\gg v_A\sim 1~cm/s$,
$\theta\sim \pi/2$.  Thus, electrons moving almost across Alfvenic layer
can absorb these waves.

Crucial and open questions for future exploration are:

a) How to enhance MAG eigen modes existing at a negligible
fluctuation level within cavities deep under bottom of the convective zone
and where is the energy source for an amplification of MAG waves in solar
interior? What is the normalization of MAG eigen modes? Full set of
ortonormalized functions (like in SHM) is still not built for these MAG
waves.

b) Does the dispersion relation that could be obtained
via a solution of kinetic equations in inhomogeneous magnetised plasma
coincide with the dispersion found from Eq. (\ref{dispersion2}) in the
MHD boundary problem?

c) Can we apply the simplest analysis given here to
another geometry of the background magnetic field $\vec{B}_0$, how to
generalize the analytic approach developed in present work in the case of
an non-uniform magnetic field $\vec{B}_0(z)$ and how to take off the
isothermal approximation $T_0 = const$? Latter step would allow us to
construct the sound speed profile $c_s(z)$ connected with the inverse
problem like in SHM.

As the necessary step for a generalization and for the checking of the
simple asymptotic solution found here an independent numerical solution of
the same initial Heun equation Eq. (\ref{master}) or even of the whole MHD
equation system Eq.  (\ref{linear}) is needed to confirm,
first, the presence of spikes shown in our Figs. 8a, 8b.

An answer on
the first question relies on the excitation of a contour through the
connection (via a non-linear interaction) with the second (neighbouring)
contour excited by the given generator.  In the oscillation theory some
mechanisms for energy transfer in two- or n-contour
systems are well-known. They are based on solutions of the telegraph
equation in electrodynamics for two (or n-cell) connected contours  or of
the non-linear oscillator equations for two (n-cell) connected pendulums
in mechanics.

The key question whether the
energy transfer time would be shorter than the collisionless damping time
of eigen modes in neighbouring contours has a promising solution for high
harmonics $n\gg 1$ for which the temporal damping occurs small in our
model, $d(n)\ll 1$.

The non-linear interaction in hydrodynamics is given
by the $\vec{v}\nabla \vec{v}$-term that was omitted in the linear theory.
This term can lead to an amplification of the mode $v_1$ if the second
mode $v_2$ (excited in a contour displaced in space with respect to the
first one) was generated by an external source.  On this way one can
imagine a turbulent source in the convective zone (with a spectrum
including low frequencies in our model ($(a~ few~ days)^{-1}$)  with the
continuous transfer of energy down the bottom of the convective zone.

One can not exclude also another energy source connected with radiative
losses and due to some spatial inhomogeneities of thermonuclear reactions
in central region or a diffusion there (for instance due to $^3$He
diffusion near $r\sim 0.2R_{\odot}$).  In that case a source $Q$= the rate
of all energy density sources and losses arises in the r.h.s. of the
energy conservation law Eq.  (\ref{pressure}).

To judge of whether MHD regular waves may be a plausible source of matter
noise let us note how they can be converted to a noise. Even infinite
sum over eigen-values (radial numbers $n=0,1..$ and angular moments $l$)
does not't lead to the noise as it is argued in \cite{Burgees} for
g-modes in SHM.

The presence of regular
waves with a small wave length is only a {\it motivation} but not {\it
realization} of the white noise. Really an instability (like buoyance force
disappearance at the bottom of the convective zone as a source of acoustic
turbulence there) can lead to the white noise with continuous spectrum due
cascade or inverse cascade processes with {\it any change of the
turbulence scale}. This is the {\it non-linear} process with the decay (or
fusion) of regular modes (of number 1 and number 2) obeying the
conservation laws: $\omega = \omega_1 + \omega_2$, $\vec{k} = \vec{k}_1 +
\vec{k}_2$. In hydrodynamics or in MHD this corresponds to the non-linear
term $(\vec{v}\nabla )\vec{v}$ omitted in linear approximation in SHM
($B_0=0$) and in linear MHD.

Thus, without any assumption about a neutrino magnetic moment we have
proposed a way to modify the MSW solution to SNP. The registration of 1-10
days variations of solar neutrino flux in underground detectors with large
statistics of neutrino events seems to be a signature of MHD wave
influence. The search of Day/Night (D/N) variations due to Earth effects
for neutrino oscillations has been already carried out in the
SuperKamiokande (SK) experiment to check the MSW scenario. It seems to be
not difficult to recognize a few days periods from the Fouirer analysis of
the SK data if MAG wave density perturbations would occur large enough
(!?).  \vskip 0.3cm

\centerline{\bf Acknowledgements}
\vskip 0.3cm
The authors thank Cliff Burgess, Alexander Rez, Jose Valle and especially
Alexei Bykov for fruitful discussions. This work has been supported by
RFFR grants 97-02-16501, 98-02-17062 and by INTAS grant 96-0659 of the
European Union.

\vskip 0.3cm

%%%%%%%%%%%%%%%%%%%%%%%%%%%%%%%%%%%%%%%%%%%%

\newpage
\centerline{\bf Figure Captions}

{\bf Fig. 1}
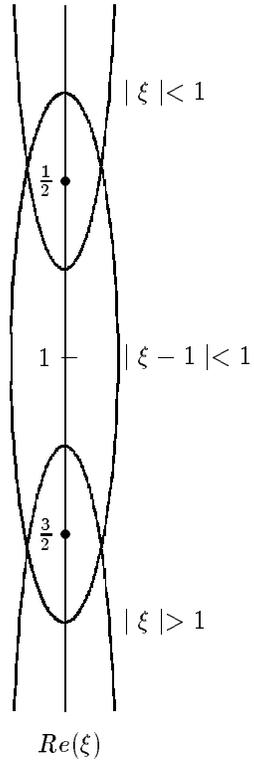
The integration region $0\leq \xi_1 =Re~\xi\leq
\xi_{1\infty}$ for all three asymptotic solutions (downwards- to the
center of the Sun). Here $\xi_1 = 1/2$ is the mathing point for the
solutions $u_2(\xi)$ and $u_3(\xi)$ and $\xi = 3/2$ is the matching point
for $u_1(\xi)$ and $u_2(\xi)$

{\bf Fig. 2.}
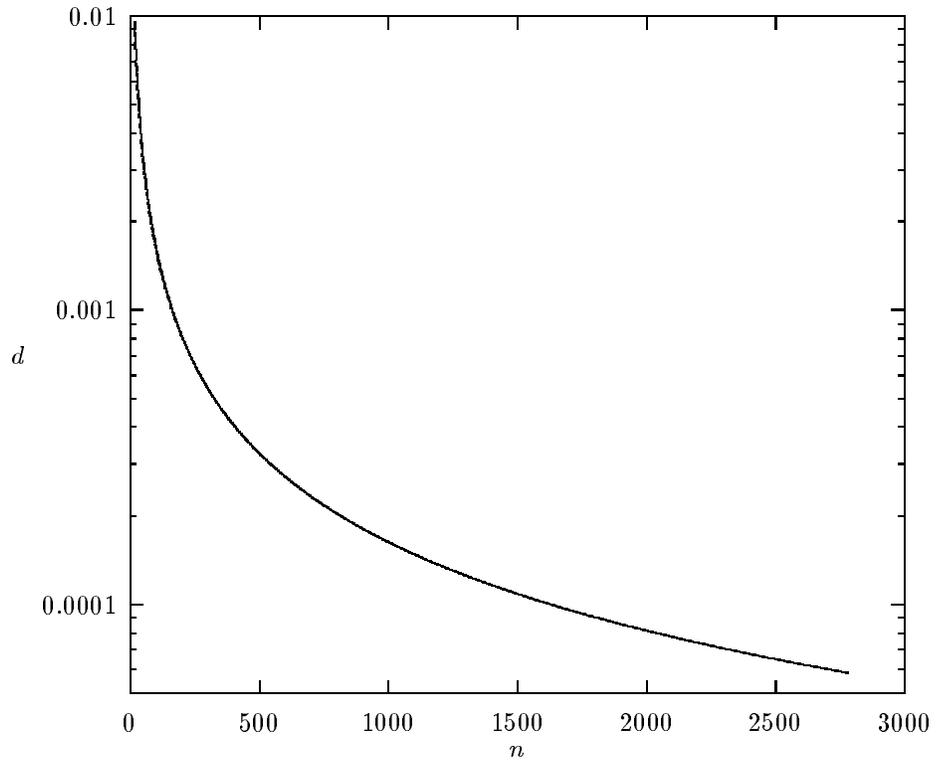
The ratio of imaginary and real parts of the eigen
frequencies $d = \mid \Omega_2\mid /\Omega_1$ in dependence on the node
number n

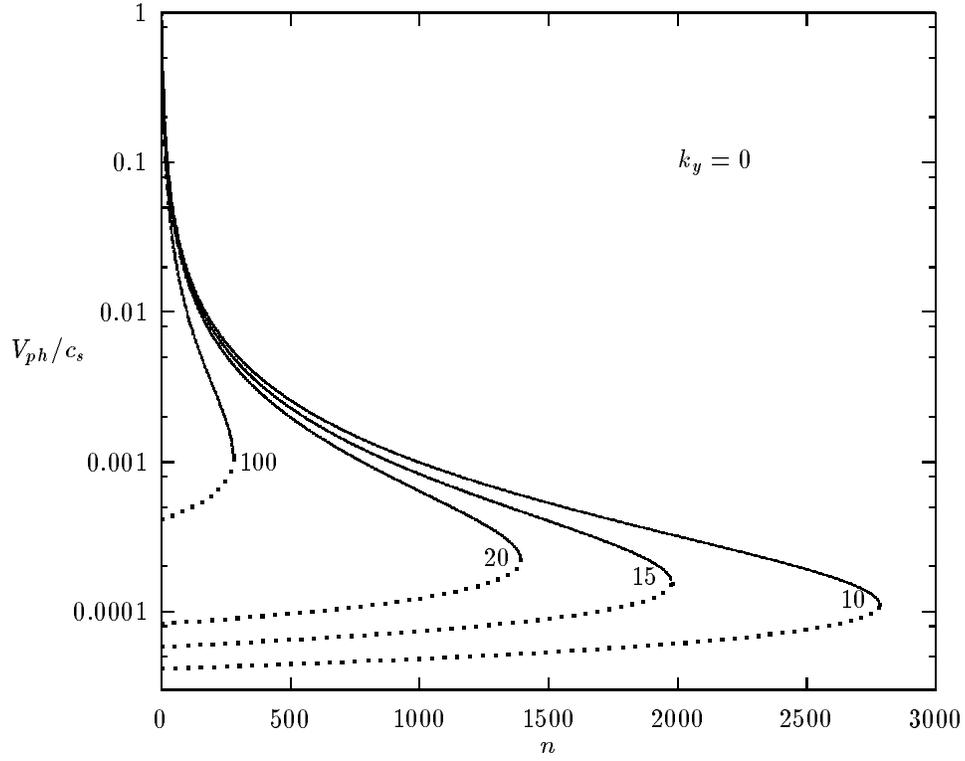
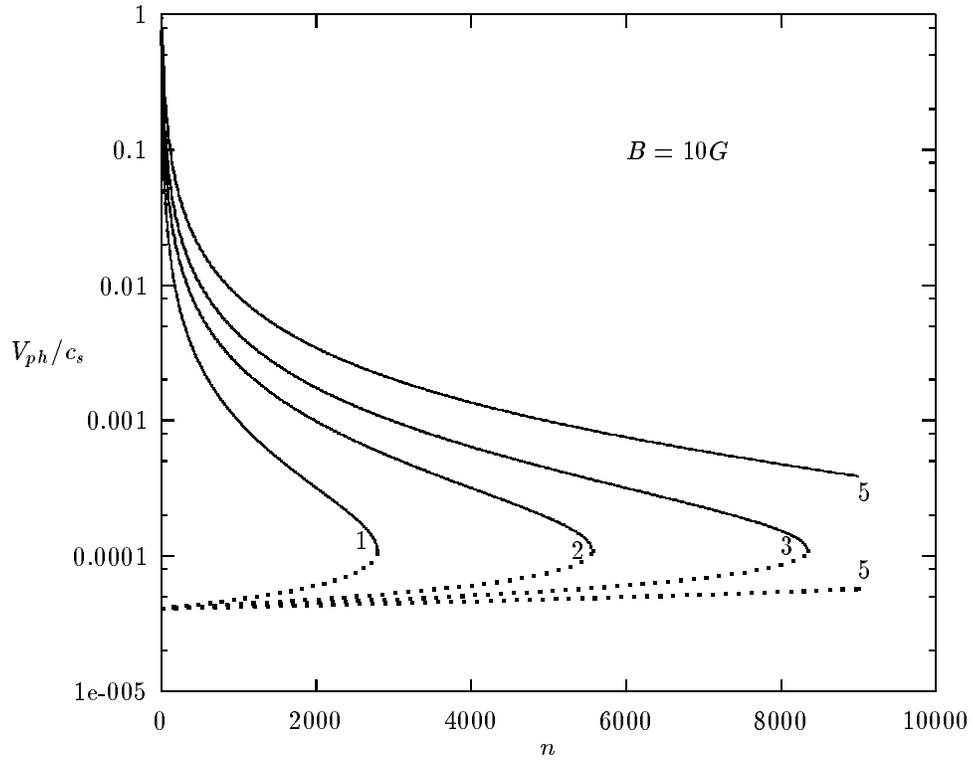
{\bf Fig. 3a and b.} The normalized phase velocity $v_{ph}/c_s = v_A/c_s$
in dependence on the node number $n$: {\bf a} for the longitudinal wave
propagation $k_y = 0$ and for different values of the ambient magnetic
field $B_0$ (in Gauss), {\bf b} for the fixed $B_0 = 10~G$ and for
different angles of the wave propagation $k_{\perp}/k_x =
1/\cos \alpha = 1,2,3,5$. The low (dashed) and high (solid) frequency
branches matched at $n_{max}$ are plotted

{\bf Fig. 4.}
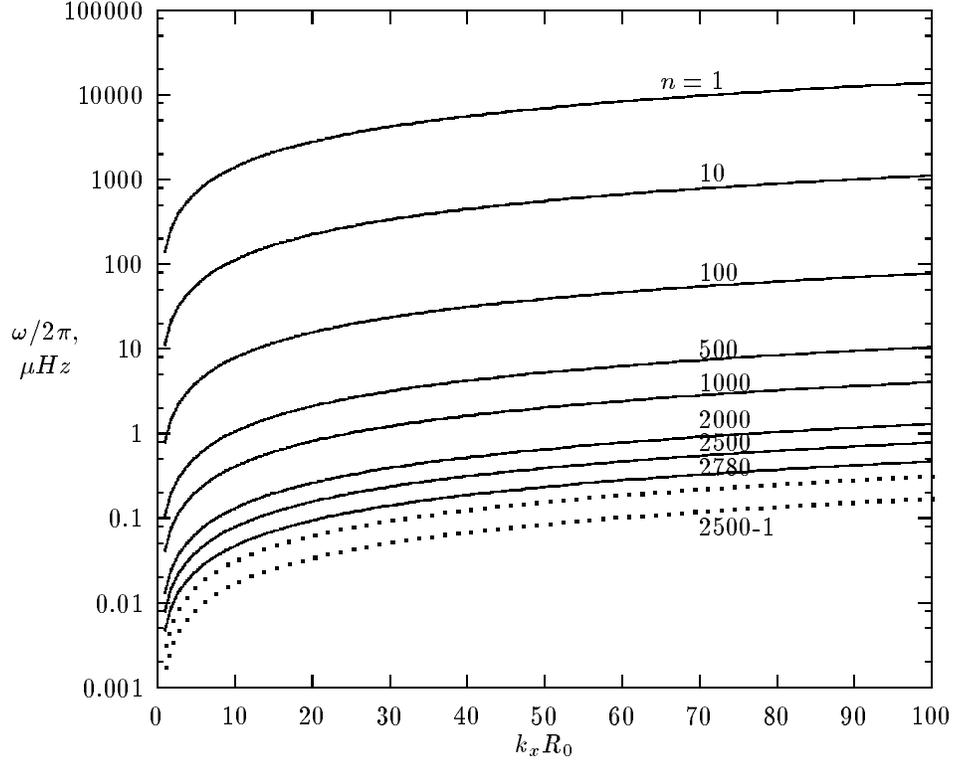
The spectrum $f = \omega_1(n,B_0, k_x)/2\pi$ for
the longitudinal wave propagation ($k_y = 0$). The ambient magnetic field
$B_0 = 10~G$ is chosen.  The low frequency branches (dashed lines)
and the high frequency ones (solid lines)  matched
at $n_{max} = 2780$ are plotted for different node numbers n.
For short wave lengths $\lambda_x = 2\pi/k_x$ the
characteristic periods of MAG waves $T = f^{-1}= (1~\div 10~\mu
Hz)^{-1} = 11.57~\div~1.157~days$ correspond to the
node numbers $n= 300\div 2000$

{\bf Fig. 5a and b.}
The Alfven layer position $z_s(n)$: {\bf a} for the
longitudinal wave propagation $k_y = 0$ in dependence on node numbers n
for the ambient magnetic field values $B_0 = 10\div 100~G$,
{\bf b} for the fixed background field $B_0 = 10~G$ in dependence on the
wave propagation angles, $(\cos \alpha)^{-1} = k_{\perp}/k_x = 1,2,3,5$.
Dashed curves respond to too low frequency branches
$T = f^{-1}\geq T_{\odot}$ beyond the approximation of the Sun at rest,
$T\ll T_{\odot} = 27~days$

{\bf Fig. 6}.
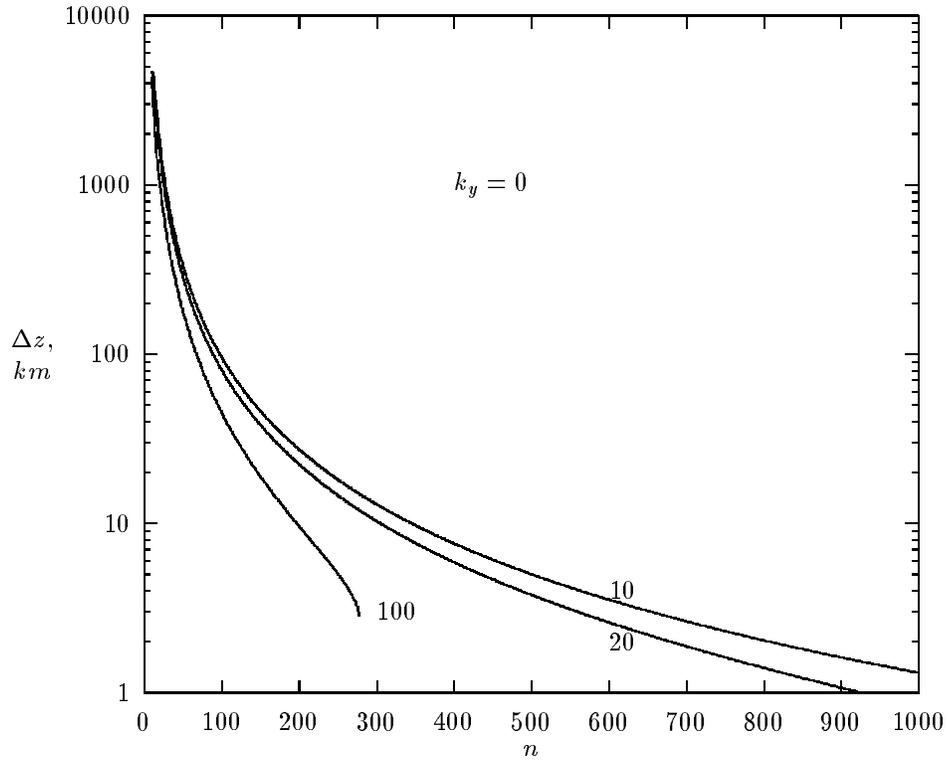
The width $\Delta z_s(n)$ of Alfven resonant layers for
the longitudinal wave propagation $k_y = 0$ in dependence on the node
number $n$ for the ambient magnetic field values $B_0 = 10,~20,~100~G$

{\bf Fig. 7a and b.}
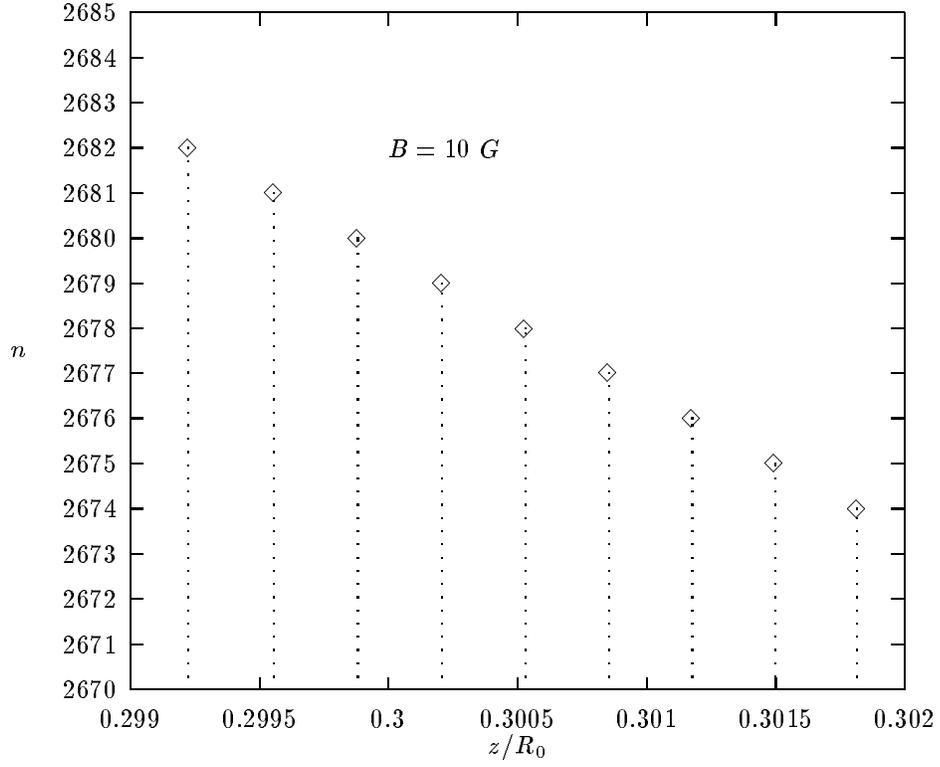
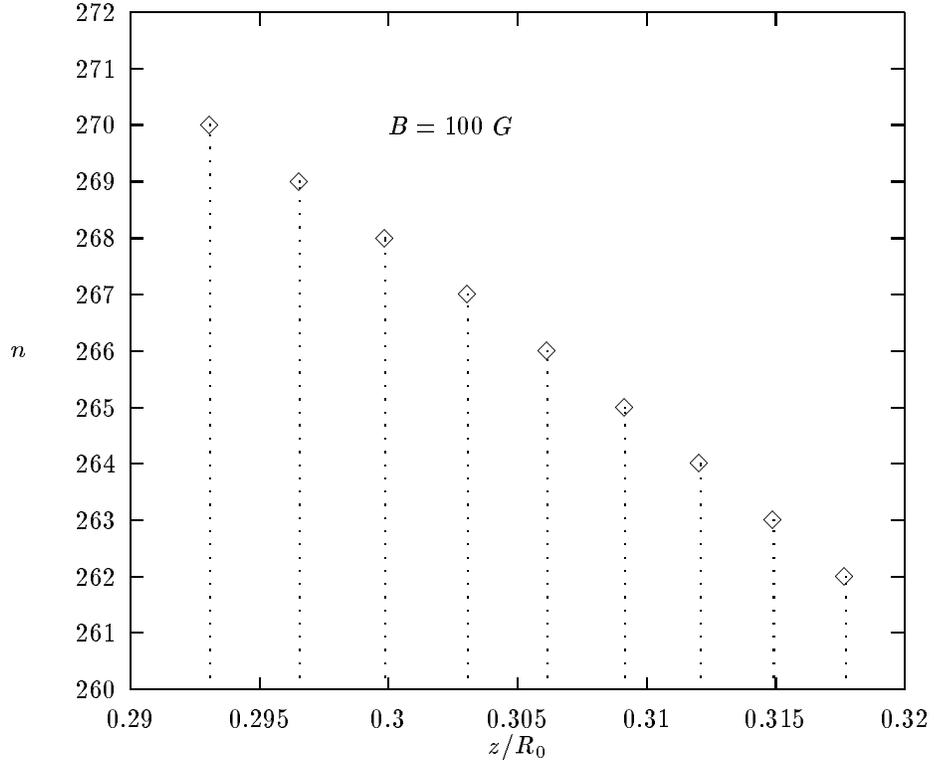
The positions $z_s(n)$ for Alfven layers near the
MSW resonant region $z_{MSW}\approx 0.3R_{\odot}$ for the ambient magnetic
field: {\bf a} $B_0 = 10~G$ (the distance between neibouring layers equals
to $z_s(n -1) - z_s(n)\sim 200~km$), {\bf b} $B_0 = 100~G$ (the distance
between neibouring layers equals to $z_s(n - 1) - z_s(n)\approx 2000~km$)

{\bf Fig. 8a and b.}
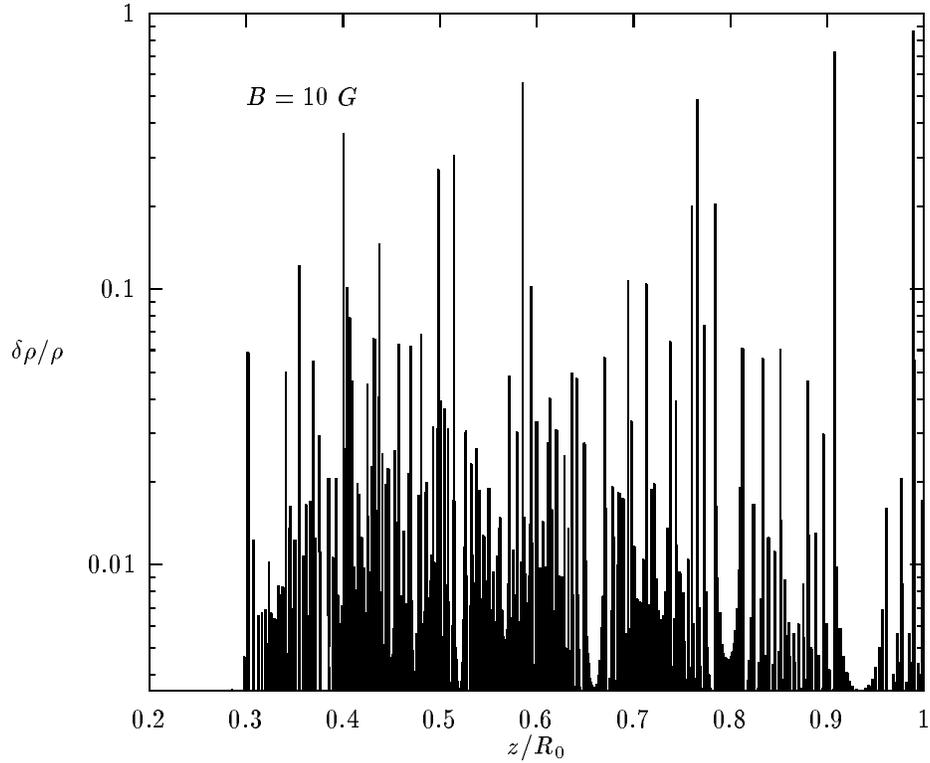
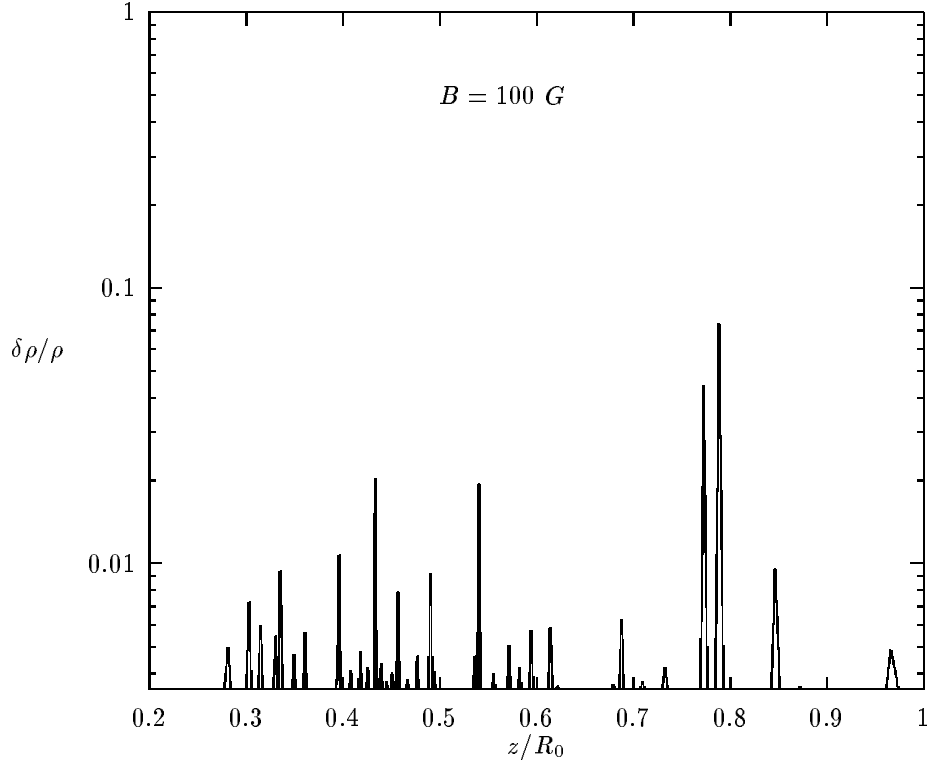
The density profile $\delta \rho(z)/\rho_0(z)$
in the case of longitudinal wave propagation $k_y = 0$. The spikes
(with unnormalized amplitudes) respond to Alfven resonant layers
with the node numbers: {\bf a} for the ambient magnetic field $B_0 = 10~G$
from $n\sim 300$ near the photosphere $z_s = R_{\odot}$ up $n = 2680$ at
the MSW region $z_s= 0.3R_{\odot}$,
{\bf b} for the ambient (large-scale) field $B_0 = 100~G$ from
$n\sim tens$ at the photosphere $z_s = R_{\odot}$ up $n = 268$
at the MSW resonant point $z_{MSW} = 0.3R_{\odot}$.
Small amplitudes of regular MAG waves that fill each cavity from
the center of the Sun up the corresponding right boundary $z_s(n)$ (up
a spike) are not seen between spikes while really the wave length within
an inner cavity, $\lambda_z = 2\pi/k_z = z_s/n$, is very
short since $n\gg 1$

{\bf Fig. 9a and b.}
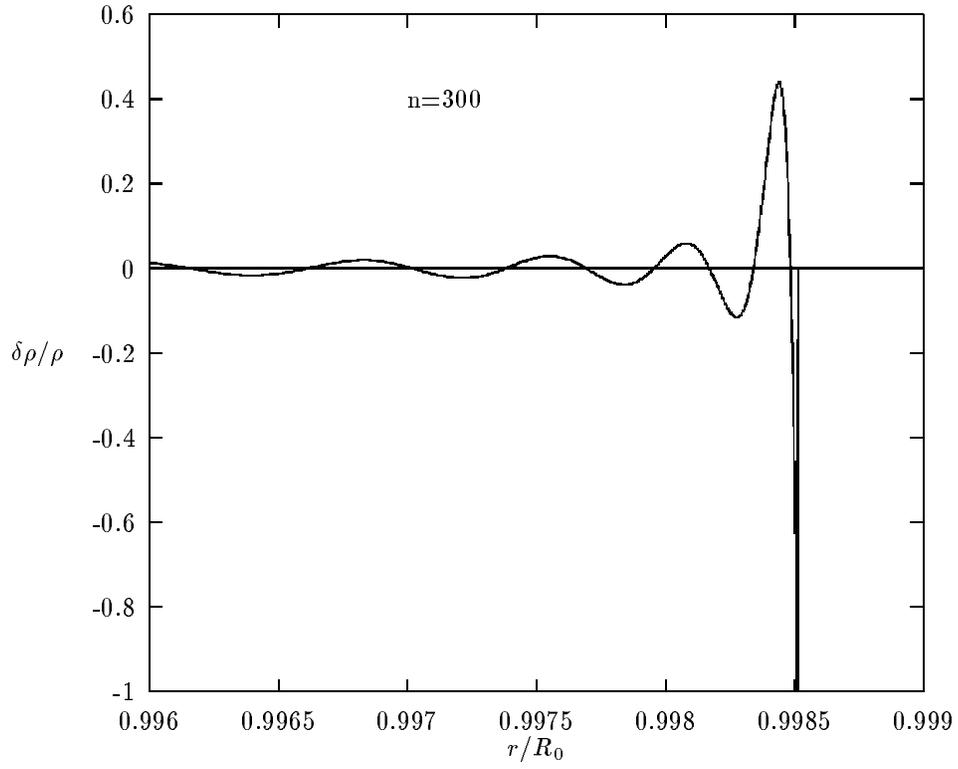
The density perturbation profile $\delta \rho_n
(z)/\rho_0$ for the wave mode $n = 300$ in the background field $B_0 =
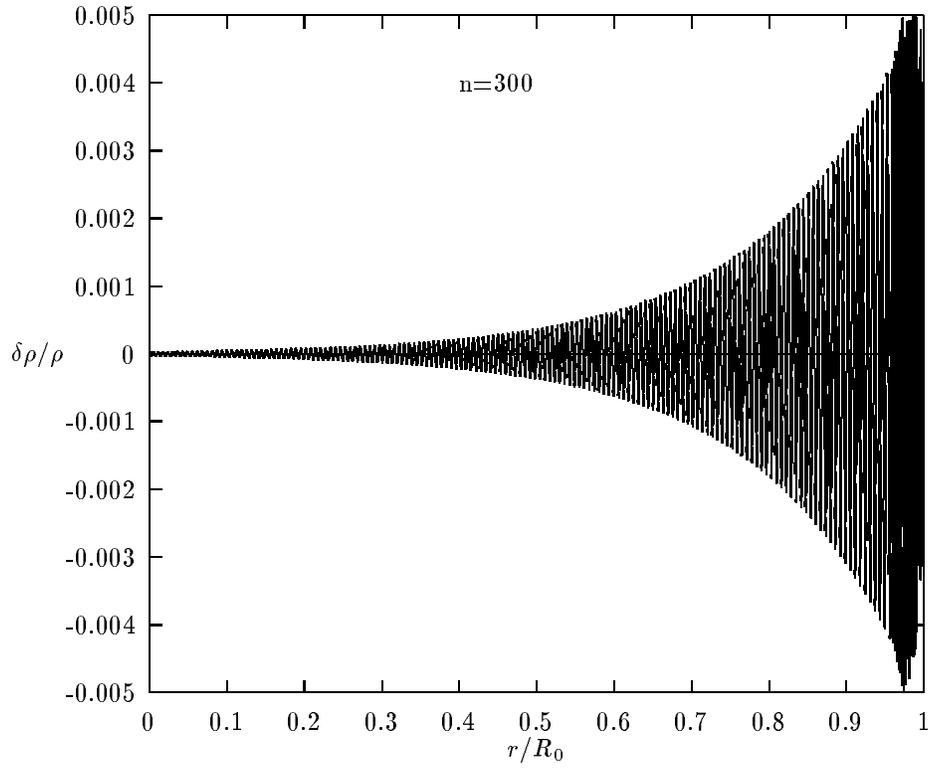
10~G$: {\bf a} small-scale structure of the density
perturbation mode n = 300 near the resonant layer position $z_s/R_{\odot} =
0.9985$, {\bf b} the same density perturbation mode within the largest
cavity with the upper bound near the photosphere, $z_s/R_{\odot}\sim 0.99$
(not seen on the large scale $0\leq z\leq R_{\odot}$),


\begin{thebibliography}{99}
%%%%%%%%%%%%%%%%%%%%%%%%%%%%%%%%%%%%%%%%%%%%%
\bibitem{Lande}
K. Lande (Homestake Collaboration) in{\em Neutrino '98}, Proceedings of
the XVIII International Conference on Neutrino Physics and Astrophysics,
Takayama, Japan, 4-9 June 1998, edited by Y. Suzuki and Y. Totsuka, to be
published in Nucl. Phys. B (Proc. Suppl.).

\bibitem{Fukuda}
Y. Fukuda et al. (Kamiokande Collaboration), Phys. Rev. Lett. 77 (1996)
1683.

\bibitem{Gavrin}
V. Gavrin (SAGE Collaboration) in Neutrino '98 [1].

\bibitem{Kirsten}
T. Kirsten (GALLEX Collaboration) in Neutrino '98 [1].

\bibitem{Suzuki}
Y. Suzuki (SuperKamiokande Collaboration) in Neutrino '98 [1].

\bibitem{Bahcall1}
J.N. Bahcall, S. Basu, M.H. Pinsonnealult, Phys. Lett. B433 (1998) 1.

\bibitem{Fiorentini}
V. Castellani, S. Degl'Innocenti, G. Fiorentini, Astron. Astrophys. 271
(1993) 601.

\bibitem{Berezinsky}
V. Berezinsky, astro-ph/9710126, invited lecture at 25th International
Cosmic Ray Conference, Durban, 28 July - 8 August, 1997.

\bibitem{vacuum}
V.N. Gribov, B.M. Pontecorvo, Phys. Lett. B28 (1969) 493.

\bibitem{MSW}
S.P. Mikheev, A.Yu. Smirnov, Sov. J. Nucl. Phys. 42 (1985) 913;\\ Nuovo
Cimento C9 (1986) 17; L. Wolfenstein, Phys. Rev. D17 (1978) 2369.

\bibitem{RSFP}
C.S. Lim, W.J. Marciano, Phys. Rev. D37 (1988) 1368; E.Kh. Akhmedov, Sov.
J. Nucl. Phys. 48 (1988) 382; Phys. Lett. B213 (1988) 64.

\bibitem{Dziembowski}
W.A. Dziembowski, Bull. Astron. Soc. India 24 (1996) 133; S.
Degl'Innocenti, W.A. Dziembowski, G. Fiorentini, B. Ricci, Astropart.
Phys. 7 (1997) 77.


\bibitem{Parker} E.N. Parker, {\em Cosmical
magnetic fields}, Clarendon Press, Oxford, 1979;\\ V.A. Kutvitskii and
L.S.  Solov'ev, Sov. Phys. JETP, 78 (1994) 456.


\bibitem{SV}
V. Semikoz, J.W.F. Valle, Nucl. Phys. B425 (1994) 651; erratum Nucl.
Phys. B485 (1997) 545 [ hep-ph/9607208];\\
 S. Sahu, V. Semikoz and J.W.F. Valle, hep-ph/9512390;\\
P. Elmfors, D. Grasso and G. Raffelt, Nucl. Phys. B479 (1996) 3;\\
J.C.D'Olivo, J.Nieves, Phys. Lett. B383 (1996) 87;\\

\bibitem{NSSV}
H. Nunokawa, V.B. Semikoz, A.Yu. Smirnov, J.W.F. Valle , Nucl. Phys.
B501 (1997) 17-40, e-print Archive: hep-ph/9701420.

\bibitem{KS}
A. Kuzenko and G. Segre, Phys. Rev. Lett. 77 (1996) 4872.

\bibitem{primeval}
N. Boruta, Ap. J. 458 (1996) 832.

\bibitem{Bahcall}
John N. Bahcall, {\em Neutrino Astrophysics}, Cambridge University
Press, 1988.

\bibitem{Krastev}
P.I. Krastev, A.Yu. Smirnov, Phys. Lett. B 338 (1989) 341; Mod. Phys.
Lett. A 6 (1991) 1001.

\bibitem{Petcov}
A. Schafer, S.E. Koonin, Phys. Lett. B185 (1987) 417;\\
R.F. Sawyer, Phys. Rev. D 42 (1990) 3908;  A. Abada, S.T. Petcov,
Phys. Lett. B 279 (1992) 153.

\bibitem{Hiroshi}
H. Nunokawa, A. Rossi, V.B. Semikoz and J.W.F. Valle, NPB 472 (1996) 495.

\bibitem{Burgees}
P. Bamert, C.P. Burgess and D. Michaud,
{\em Neutrino Propagation Through Heliosesmic Waves},
Preprint McGill-97/13, hep-ph/9707542;\\
C.P. Burgess, D. Michaud, Ann. Phys., (NY) 256 (1997) 1.

\bibitem{Balantekin}
F.N. Loreti, A.B. Balantekin,  Phys. Rev. D50 (1994) 4762.

\bibitem{Zhugzhda} Yu.D. Zhugzhda, N.S.  Dzhalilov,
Geophys.  Astrophys. Fluid. Dynamics, 35 (1986) 131.


\bibitem{Guzzo}
M.M. Guzzo, N. Reggiani and J.N. Colonia, Phys, Rev. D56 (1997) 588.

\bibitem{Stix}
M. Stix, {\em The Sun}, Springer-Verlag (Berlin, New York, London,
Paris, Tokio) 1989, 890 pp.


\bibitem{Dzhalilov2}
N.S. Dzhalilov, Yu.D. Zhugzhda,
Astron. Zh., 67 (1990) 561 (in Russian, translated to English).

\bibitem{Heun}
K. Heun, Math. Ann. 33 (1889) 161-179.


\bibitem{Priest2}
E.R. Priest, {\em Solar Magnetohydrodynamics}, by D. Reidel Publishing
Company, Dordrecht, Netherlands, 1982.

\bibitem{Ionson}
J. Ionson, Ap. J. 226 (1978) 650; 254 (1982) 318; 276 (1984) 357;\\
J.V. Hollweg, Ap. J. 312 (1987) 880; 320 (1987) 875.

\bibitem{Hasegawa}
A. Hasegawa, C. Uberoi, {\em The Alfven wave}, Washington D.C.,
Technical   Inform. Center, UUS Depart. of Energy, 1988.


\bibitem{Priest}
L. Nocera, B. Leroy, E.R. Priest, Astron. Astrophys., 133 (1984)  387.


\bibitem{Dzhalilov}
V.N. Oraevsky, N.S. Dzhalilov, Astron. Zh. 74 (1997) 99 (in Russian);\\
Translated in: V.N. Oraevskii, N.S. Dzhalilov,
Astronomy Reports, 41 (1997) 91.


\end{thebibliography}
\end{document}